\documentclass[preprint2]{emulateapj}

\pdfoutput=1
\usepackage{apjfonts}
\usepackage{multirow}
\usepackage{amssymb}
\usepackage[fleqn]{amsmath}
\usepackage{graphicx}
\usepackage[usenames,dvipsnames]{color}
\usepackage[colorlinks=true, linkcolor=BrickRed, citecolor=Blue, urlcolor=Blue, filecolor=Blue]{hyperref}

\slugcomment{To appear in ApJ}

\shorttitle{Dark stars, reionization and the CMB}
\shortauthors{Scott et al.}

\begin{document}

\title{Impacts of Dark Stars on Reionization and Signatures in the Cosmic Microwave Background}

\author{Pat Scott\altaffilmark{1}$^*$, Aparna Venkatesan\altaffilmark{2}$^\dagger$, Elinore Roebber\altaffilmark{1}, Paolo Gondolo\altaffilmark{3}, Elena Pierpaoli\altaffilmark{4} \& Gil Holder\altaffilmark{1}}
\altaffiltext{*}{E-mail: patscott@physics.mcgill.ca}
\altaffiltext{$\dagger$}{E-mail: avenkatesan@usfca.edu}
\altaffiltext{1}{Department of Physics, McGill University, 3600 rue University, Montreal, QC H3A 2T8, Canada}
\altaffiltext{2}{Department of Physics and Astronomy, University of San Francisco, 2130 Fulton St. San Francisco, CA 94117, USA}
\altaffiltext{3}{Department of Physics and Astronomy, University of Utah, Salt Lake City, UT 84112, USA}
\altaffiltext{4}{Department of Physics and Astronomy, University of Southern California, Los Angeles, CA 90089, USA}

\begin{abstract}

We perform a detailed and systematic investigation of the possible impacts of dark stars upon the reionization history of the Universe, and its signatures in the cosmic microwave background (CMB).  We compute hydrogen reionization histories, CMB optical depths and anisotropy power spectra for a range of stellar populations including dark stars.  If dark stars capture large amounts of dark matter via nuclear scattering, reionization can be substantially delayed, leading to decreases in the integrated optical depth to last scattering and large-scale power in the EE polarization power spectrum.  Using the integrated optical depth observed by WMAP7, in our canonical reionization model we rule out the section of parameter space where dark stars with high scattering-induced capture rates tie up $\ga 90\%$ of all the first star-forming baryons, and live for $\ga250$\,Myr.  When nuclear scattering delivers only moderate amounts of dark matter, reionization can instead be sped up slightly, modestly increasing the CMB optical depth.  If dark stars do not obtain any dark matter via nuclear scattering, effects upon reionization and the CMB are negligible.  The effects of dark stars upon reionization and its CMB markers can be largely mimicked or compensated for by changes in the existing parameters of reionization models, making dark stars difficult to disentangle from astrophysical uncertainties, but also widening the range of standard parameters in reionization models that can be made consistent with observations.

\end{abstract}


\section{Introduction}
\label{intro}

At a redshift of $z\sim 1100$, the bulk of electrons and protons in the Universe recombined to form neutral hydrogen.  This cleared the way for thermal photons to free-stream away from the resulting surface of last scattering, forming what we now see as the cosmic microwave background (CMB; see e.g.~\citealt{HuDodelson,CMBpedestrian}).  A prolonged period of darkness ensued, until the first sources of hard ionizing radiation forming within galaxies appeared at redshifts $z\lesssim30$ \citep{Gnedin:00, Ciardi:00,Bromm:01,Wyithe:03,Schaerer02,Venkatesan:03,Benson:06,Loeb:09}.  These sources are thought to have reionized the neutral gas in the intergalactic medium (IGM).  This process was completed by $z\sim$ 7 \citep{Fan:06,Dawson:07}, with the IGM subsequently remaining in the fully-ionized state we see it in today.  

The process of reionization is expected to have left its imprint upon a number of cosmological observables.  The CMB is sensitive to the total optical depth to the surface of last scattering; any free electrons between us and the surface will have scattered CMB photons before they could reach us, modifying the anisotropy power spectra we observe today.  Similarly, the changing distribution of neutral hydrogen during reionization can be mapped using its ground-state hyperfine transition \citep{Furlanetto:06,Chen:08}, corresponding to a rest-frame wavelength of 21\,cm.  The signal is expected to be weak because this is a forbidden line, so the transition probability is small.  A multitude of upcoming experiments hope to detect it nonetheless \citep{Morales:10}; these include the Low-Frequency (LOFAR; \citealt{Harker10}), Murchison Wide-Field (MWA; \citealt{MWA}) and Square Kilometer Arrays (SKA; \citealt{Carilli08}).

The first ionizing sources are generally thought to be the first stars (see, e.g., \citealt{Gnedin:00,Bromm:01,Venkatesan:03,Tumlinson:04,Wise:08}), referred to as Pop~III. 
The terms Pop~III and Pop~II are typically interpreted as broad distinctions based on composition - respectively, they represent stars that are metal-free and metal-poor.
Pop~III stars consist entirely of primordial hydrogen and helium synthesized in the Big Bang, so are likely to have a mass function weighted towards higher stellar masses than that of Pop~II, due to the absence of metal lines, which allow efficient gas cooling and cloud fragmentation in metal-enriched galaxies at later epochs \citep{Bromm:01,Tumlinson:03}.
Within the Pop~III category, the terms Pop~III.1 and Pop~III.2 have recently arisen in the literature. Typically, Pop~III.1 connotes the {\it very} first stars (i.e., first-generation metal-free stars) that have masses exceeding $\sim$$100-300\,M_\odot$, and that form in $\sim$$10^6\, M_\odot$ dark matter (DM) minihalos.  Pop~III.2 stars form in the wake of the radiative and chemodynamic feedback of Pop~III.1 supernovae.  Pop~III.2 are therefore ``second-generation'' stars with metal-free composition, and are thought to have lower stellar masses  on average ($\sim$$10-100\,M_\odot$) than Pop~III.1 stars (\citealt{Tumlinson:04,McKee:08,Ohkubo:09}; see, however, \citealt{Clark:11} for arguments on Pop III.1 stars having {\it lower} stellar mases than Pop~III.2 stars).
Although Pop~III.1 stars were initially thought to form in isolation, producing a single very massive star per minihalo \citep{Abel:02}, this paradigm has lately given way to one where they form mostly in pairs, or systems of even higher multiplicity \citep{Krumholtz:09, Turk:09, Stacy:10}.

Recent observations of the galaxy luminosity function (LF) at high redshifts, $z=4-8$ \citep{Bouwens:10, Bouwens:11} indicate that the earliest, faintest galaxy halos make a substantial contribution to reionization, in good agreement with the hypothesis that the first stars are indeed the first ionizing sources. Connecting stellar population models to UV observations of the faint-end galaxy LF is fraught with systematic uncertainties, however, as it requires extrapolation of the observations to even higher redshifts ($z \ga$ 8), and making a number of assumptions with regard to the parameterization of the underlying reionization models. For these reasons, and given the relative insensitivity of present-day 21 cm data to reionization physics, we focus on constraints from the CMB in this paper.

It has been shown \citep{Spolyar08, Natarajan08b} that dark matter could have had a more direct impact on Pop~III.1 star formation than simply facilitating the initial baryonic collapse.  If DM consists of a new particle that is present in equal numbers to its antiparticle, or if it is indeed its own antiparticle, it will self-annihilate to produce Standard Model (SM) particles such as quarks, photons and electrons.  This is the case for e.g.~weakly-interacting massive particles (WIMPs), arguably the most widely-studied and natural solution to the dark matter problem \citep[see e.g.][]{Jungman96, Bergstrom00, Bertone05, Bertonebook}.  As baryons cool and contract during star formation, they steepen the gravitational potential within the minihalo, drawing even more dark matter into its center \citep{Freese08b}.  The resultant spike in the DM annihilation rate can inject an appreciable amount of energy into the collapsing cloud, halting or delaying star formation \citep{Mapelli06,Stasielak06,Ripamonti07b,Ripamonti10} and resulting in a cool, partially-collapsed object known as a ``dark star'' \citep{Spolyar08,Natarajan08b}.  Ongoing annihilation of DM particles in the core of a star can have substantial impacts upon its structure and evolution \citep{SalatiSilk89,Fairbairn08,Scott08a,Iocco08a,Iocco08b,Freese08c,Scott09,Spolyar09}.

As dark stars have significantly different structures and evolutionary histories to ``normal'' Pop~III stars, their ionizing photon outputs differ substantially as well \citep{Yoon08}, leading to a potentially distinct impact upon the process of reionization \citep{Schleicher08}.  The purpose of this paper is to systematically investigate the effects of dark stars upon the reionization history of the Universe.  We begin by discussing dark star formation and evolution in Sec.~\ref{darkstars}, and stellar population models including dark stars in Sec.~\ref{darkpop}.  We give an overview of our reionization models and calculations in Sec.~\ref{reionization}.  In Sec.~\ref{reionhist} we give the resulting alternative reionization histories for Universes containing dark stellar populations.  In Sec.~\ref{cmb}, we show how such IGM ionization histories would impact the measured optical depth to electron scattering in the CMB, drawing limits on dark star populations using the integrated optical depth measured by WMAP7, and making predictions for the corresponding constraining power of Planck.  We also show the impact of dark stars upon the EE polarization anisotropy spectrum of the CMB, and discuss its potential use for constraining dark star populations. We conclude in Sec.~\ref{conclusion}.

\section{Formation and evolution of dark stars}
\label{darkstars}

Assuming DM self-annihilates, the formation of a dark star relies on some efficient means for bringing the DM into the center of a star.  The two processes which may provide this means are gravitational contraction and nuclear scattering.  

The first is simply an effect of the changing gravitational potential during the collapse of a baryonic gas cloud \citep{Freese08a}.  As baryons cool and collapse onto the central overdensity, dissipating energy by radiative emission and angular momentum by hydrodynamic and magnetic interactions \citep[which may be enhanced during the formation of the first stars by dynamo effects;][]{Schleicher10,Sur10}, the gravitational potential in the core of the halo steepens.  In turn, the steepening potential draws DM into the center of the cloud (despite its inability to actually dissipate energy), resulting in a strongly-peaked DM density distribution.  This is referred to simply as `gravitational contraction,' and is more general than the well-known case of adiabatic contraction because it does not strictly require that the gravitational potential change more slowly than the orbital timescale of individual particles.  In the canonical scenario, where a single Pop~III.1 star forms at the center of the very first halos, stars begin their lives already in possession of a large reservoir of DM.  

The second way for DM to end up in a stellar core relies on it possessing a weak-scale scattering cross-section with nucleons.  Such an interaction is characteristic of WIMP dark matter.  Assuming such a cross-section, DM particles passing through a star can lose energy through collisions with stellar nuclei, becoming gravitationally bound to the star \citep{Steigman78,Krauss85,Press85,Gould87b}.  This leads to repeat scattering events, eventually removing enough energy that the particle ends up in the stellar core.

Regardless of the path DM follows into a star, the effects are essentially the same.  DM annihilation in the core provides an additional energy source alongside nuclear fusion, causing the core to expand and cool.  This occurs due to the negative specific heat of a self-gravitating body, and the fact that the DM annihilation rate is decoupled from the nuclear core density.  The core expansion leads to a larger, cooler, typically strongly convective stellar object \citep{Spolyar08,Scott09,Spolyar09,Casanellas09}.  In the case of gravitationally-contracted DM, the collapse of the forming star is slowed, effectively extending the protostellar phase.  The slowdown allows gas accretion to continue longer than it otherwise would, as the onset of radiative feedback is delayed.  The resulting object therefore grows more than in the absence of DM \citep{Spolyar08,Umeda09}, leading to masses of order $\sim$800--1000\,$M_\odot$.  It has been suggested that supermassive objects might even be possible \citep{FreeseSMDS}, though this is strongly constrained by existing data \citep{Zackrisson10b}.

The degree to which a star affected by DM annihilation resembles either a fully-fledged dark star (during the extended protostellar phase) or a main-sequence object depends upon the rate of annihilation in its core.  Higher rates of annihilation are required to support larger, more diffuse, protostellar-like structures against further collapse.  The evolution of a dark star therefore depends strongly on the rate at which dark matter is delivered to the stellar core, and how that rate changes over time.  Typically, gravitationally-contracted dark matter will be exhausted in a period of $\sim$0.4\,Myr \citep{Spolyar09}.\footnote{We note however that some uncertainty remains over this value, with much shorter timescales suggested by \citet{Iocco08b}.  \citet{Ripamonti10} even show that at least in the early stages of the collapse, dark matter annihilation might in fact \textit{help} the gas to cool and contract rather than hinder it, by enhancing the formation of H$_2$ molecules; this effect has yet to be taken into account in most dark star modeling.}  Without replenishment via DM capture due to nuclear scattering, dark stars then contract, heat up and move on to the main sequence (MS) to live their lives as extremely massive Pop~III.1 stars.  

In the simplest scenario, where the DM halo is homogeneous and spherically symmetric, DM capture by nuclear scattering has been shown not to substantially extend the lifetime of dark stars \citep{Sivertsson10}.  The impact of more realistic halo distributions upon the DM star-crossing rate, and therefore the capture rate, remains to be understood however (see e.g.~\citealt{FreeseSMDS} for the suggestion that particle orbits may even be strongly centrophilic).  In principle, if capture rates due to nuclear scattering are high enough, dark stars may exist as cool, diffuse objects for up to $\sim$500\,Myr \citep[see e.g.][for a discussion of possible dark star lifetimes]{Zackrisson10a}.

Similarly, the impact of Pop~III.1 stars forming as binaries or higher-multiplicity systems is not yet well understood.  At some level the displacement of the collapsing baryonic core(s) from the central DM spike may indeed prevent dark star formation all together.  Alternatively, the fragmentation process may introduce sufficient structure to the phase-space distribution of the DM halo that the findings of \citet{Sivertsson10} are circumvented, and dark star lifetimes in fact become longer.  In this case however, the resulting dark stars might be of substantially lower mass than those formed in isolation, due to reduced accretion onto the central object \citep{Peters10}, shown to be significant for first star formation \citep{Clark:11}.  Further review of the structure and evolution of dark stars and the mechanisms for fueling them can be found in \citet{ScottCRF}.

Given the substantial theoretical uncertainty in predicting the longevity of dark stars, for the purposes of this paper we consider their lifetimes to be a free parameter, to be constrained by CMB data or other observations of reionization.  Although we choose a halo mass threshold at $z=20$ that is designed to allow the formation of a single dark star in the very smallest halos (see Sec.~\ref{reionization}), we do not make any strong assumptions as to the number of dark stars forming in each halo in general, as the reionization formalism described in Sec.~\ref{reionization} is non-specific as to the multiplicity of star formation.

\section{Dark and semi-dark stellar populations}
\label{darkpop}

\subsection{Population models}
\label{popmodels}

Modeling a stellar population containing dark stars requires careful consideration of the possible evolutionary histories of the dark component.  We define different structural forms for dark stars, depending upon the degree to which they are dominated by dark matter: the dark star proper (DSP) and the dark star near the main sequence (DSNMS).  The DSP structure occurs when dark matter annihilation contributes a substantial fraction of the star's energy budget, leading to the large, diffuse, cool objects discussed by \citet{Spolyar08}.  These objects lie far to the right of the HR diagram, in the region populated by protostars on their way to the main sequence.  The DSNMS structure occurs when dark matter contributes only a small amount of the star's total energy budget, and much more closely resembles a main sequence star.  Correspondingly, they lie only slightly to the right of the standard Pop~III main sequence on the HR diagram.

Of course, the two structures are not entirely distinct; a continuum of stellar structures is possible, parameterized by the amount of energy produced by dark matter annihilation in the stellar core (as discussed in Sec.~\ref{darkstars}).  Exactly what structure a dark star exhibits depends upon its age (with the DSNMS stage always following the DSP stage) and the rate at which dark matter is captured and converted into heat.  Sustained high rates of capture therefore effectively increase the duration of the DSP phase, but not the DSNMS phase, as the star is supported by DM annihilation in a cool, diffuse configuration until the DM runs out.  Correspondingly, sustained moderate capture rates increase the duration of the DSNMS phase, but not the DSP phase, as the star is supported by DM annihilation and nuclear burning, in a configuration that is only slightly cooler than the equivalent main sequence structure \citep[see e.g.][]{Scott09}.

We take a phenomenological definition for the exact demarcation between the DSP and DSNMS phases.  The DSP phase includes all structures where capture rates are sufficiently large to keep the star too cool to contribute to reionization ($Q\sim0$; cooler stars are redder, so more of their luminosity is output at wavelengths longward of the neutral hydrogen ionization threshold).  Structures in the DSNMS phase, on the other hand, have small enough WIMP capture rates to be sufficiently hot to make some contribution to reionization.  Because the ionizing photon flux for a given stellar mass falls off very abruptly as the relative contribution of dark matter annihilation to the star's total energy budget is increased, to a first approximation stars in this phase can be modeled as main-sequence objects.

\begin{deluxetable}{l r r r r r}
\centering
\tablecaption{Scenarios, parameters, redshifts of H~\textsc{i} reionization and integrated optical depths to last scattering for stellar populations with dark stars. Optical depths include contributions from He~\textsc{ii}, He~\textsc{iii} and residual electron fraction after recombination. See text for details. 
\label{bigtable}}
\scriptsize
\tablewidth{0pt}
\tablehead{
\colhead{Scenario} & \colhead{$t_\mathrm{DSP}$} & \colhead{$t_\mathrm{DSNMS}$} & \colhead{$f_\mathrm{DS}$} & \colhead{$z_\mathrm{reion}$} & \colhead{$\tau_\mathrm{e}$}\\
\colhead{} & \colhead{(Myr)} & \colhead{(Myr)} & \colhead{} & \colhead{} & \colhead{}}
\startdata
Pop~II+III only	& 0\vspace{2mm}	& 0	& 0	& 11.050	& 0.1084 \\
NC 		& 0.4 	& 1 	& 0.01 	& 11.050	& 0.1084 \\
(no capture)\vspace{2mm}&&	& 1	& 11.050	& 0.1084 \\
MC		& 0.4	& 3	& 0.01	& 11.050	& 0.1084 \\
(meager capture)&	&	& 0.1	& 11.061	& 0.1085 \\
		&	&	& 0.6	& 11.092	& 0.1089 \\
		&	&	& 1	& 11.122	& 0.1093 \\
		&	& 6	& 0.01	& 11.050	& 0.1084 \\
		&	&	& 0.1	& 11.071	& 0.1086 \\
		&	&	& 0.3	& 11.102	& 0.1091 \\
		&	&	& 0.6	& 11.143	& 0.1097 \\
		&	&	& 0.8	& 11.174	& 0.1101 \\
\vspace{2mm}	&	&	& 1	& 11.205	& 0.1105 \\
EC		& 5	& 1	& 0.01	& 11.050	& 0.1084 \\
(extreme capture)& 	& 	& 0.1	& 11.030	& 0.1080 \\
		& 	& 	& 0.3	& 10.978	& 0.1072 \\
		& 	& 	& 0.6	& 10.885	& 0.1060 \\
		& 	& 	& 0.8	& 10.822	& 0.1051 \\
		& 	& 	& 1	& 10.760	& 0.1042 \\
		& 15	& 1	& 0.01	& 11.050	& 0.1083 \\
		& 	& 	&  0.1	& 10.999	& 0.1076 \\
		& 	& 	&  0.3	& 10.885	& 0.1059 \\
		& 	& 	&  0.6	& 10.697	& 0.1031 \\
		& 	& 	&  0.8	& 10.560	& 0.1011 \\
		& 	& 	&  0.9	& 10.486	& 0.0999 \\
		& 	& 	&  1	& 10.423	& 0.0988 \\
		& 50	& 1	&  0.01	& 11.040	& 0.1083 \\
		& 	& 	&  0.1	& 10.968	& 0.1072 \\
		& 	& 	&  0.3	& 10.780	& 0.1046 \\
		& 	& 	&  0.6	& 10.476	& 0.1000 \\
		& 	& 	&  0.8	& 10.231	& 0.0962 \\
		& 	& 	&  0.9	& 10.103	& 0.0941 \\
		& 	& 	&  1	&  9.973	& 0.0917 \\
		& 150	& 1	& 0.01	& 11.040	& 0.1083 \\
		& 	& 	& 0.1	& 10.936	& 0.1069 \\
		& 	& 	& 0.3	& 10.644	& 0.1033 \\
		& 	& 	& 0.6	& 10.060	& 0.0957 \\
		& 	& 	& 0.8	&  9.514	& 0.0878 \\
		& 	& 	& 0.9	&  9.192	& 0.0824 \\
		& 	& 	& 1	&  8.852	& 0.0755 \\
		& 500	& 1	& 0.01	& 11.040	& 0.1083 \\
		& 	& 	& 0.1	& 10.926	& 0.1069 \\
		& 	& 	& 0.3	& 10.623	& 0.1032 \\
		& 	& 	& 0.4	& 10.433	& 0.1010 \\
		& 	& 	& 0.5	& 10.210	& 0.0983 \\
		& 	& 	& 0.6	&  9.941	& 0.0951 \\
		& 	& 	& 0.7	&  9.592	& 0.0910 \\
		& 	& 	& 0.8	&  9.079	& 0.0852 \\
		& 	& 	& 0.85	&  8.726	& 0.0811 \\
		& 	& 	& 0.9	&  8.215	& 0.0755 \\
		& 	& 	& 1	&  6.282	& 0.0446
\enddata
\end{deluxetable}

We define three distinct capture scenarios: no capture (NC), meager capture (MC) and extreme capture (EC).  NC is the canonical scenario discussed and simulated using a simple varying-index polytropic model by \citet{Spolyar09}: dark stars form by gravitational collapse, and grow to exhibit larger masses than typical Pop III.1 stars because annihilation of gravitationally-contracted dark matter in their cores inhibits the collapse.  After the initial (gravitationally-contracted) population of dark matter is exhausted, the star finishes contracting and makes its way to the main sequence, where it lives like any other $\sim$800\,$M_\odot$ star.  In this case, we have a DSP phase with a duration approximately equal to the ages of the stars of \citet{Spolyar09} upon reaching the MS ($t_\mathrm{DSP}\sim0.4$\,Myr for a DM mass of $m_\chi=100$\,GeV), followed by a DSNMS phase of duration equal to the standard MS lifetime of an $\sim$800\,$M_\odot$ MS Pop III.1 star ($t_\mathrm{DSNMS}\sim1$\,Myr; e.g.~\citealt{Schaerer02}). 

In the MC scenario, we have enough dark matter capture to extend the lifetime $t_\mathrm{DSNMS}$ of the DSNMS phase, whereas in the EC scenario we have enough dark matter capture to instead extend $t_\mathrm{DSP}$, the lifetime of the DSP phase.  In these two scenarios, we therefore take the lifetimes of the respective phases as free parameters, and keep the lifetime of the other phase fixed at its canonical value in the NC scenario.

In general, the maximum duration of any phase of the lifetime of a star powered by dark matter capture (either wholly or partially) is limited by the total core hydrogen-burning lifetime of the star, and the self-annihilation time of the dark matter halo from which the star captures its dark matter.  For stars in the most dark-matter dominated parts of the DSP phase, the low core temperature and very high capture rate mean that the self-annihilation time is the relevant limit, and the hydrogen-burning time plays little role.  For stars in the DSNMS phase, the opposite is true.  These limits were calculated and discussed in detail by \citet{Zackrisson10a}.  Here we approximately adopt the values $t_\mathrm{max}$ of Zackrisson et al.~as the upper limits for the durations of the two phases. The lower limits are given by the NC case, which can be seen as a limiting case of both the MC and EC scenarios (although physically, it is of course only continuously linked to the MC case).  This gives $0.4\,\mathrm{Myr} \le t_\mathrm{DSP} \lesssim 500$\,Myr and $1\,\mathrm{Myr} \le t_\mathrm{DSNMS} \lesssim 6$\,Myr.

Finally, we also have the dark star mass fraction $f_\mathrm{DS}$ as an additional free parameter in all three scenarios (NC, MC and EC).  This describes the fraction of the star-forming baryonic mass that initially goes into dark stars rather than normal Pop III stars.  We begin with an initial population consisting of a fraction $f_\mathrm{DS}$ of dark stars in the DSP phase and a complementary fraction $(1-f_\mathrm{DS})$ of normal Pop III stars.  As the normal Pop III stars finish their starburst (i.e.~after $t_\mathrm{Pop\,III}$, which we set to 10\,Myr), they are replaced with Pop II stars.  As the dark stars transition from DSP to DSNMS and then eventually die, they are replaced with either normal Pop III stars if the death occurs before the end of the original Pop III starburst, i.e.~$t_\mathrm{DSP}+t_\mathrm{DSNMS}<t_\mathrm{Pop\,III}=10$\,Myr, or directly with Pop II stars if the death occurs after the cessation of the original Pop III starburst, i.e.~$t_\mathrm{DSP}+t_\mathrm{DSNMS}\ge t_\mathrm{Pop\,III}=10$\,Myr.  In the case where the dark stars die and are replaced by a new population of Pop~III stars, these Pop~III stars later also die and are replaced by Pop II stars at $t=t_\mathrm{Pop\,III}=10$\,Myr after the beginning of star formation, just like their counterparts in the complementary fraction $(1-f_\mathrm{DS})$ of normal Pop III stars present from the beginning of the calculation.

For ease of reference, the full set of scenarios and parameters with which we compute reionization histories is given in Table~\ref{bigtable}.

For some combinations of parameters (large $f_\mathrm{DS}$ and long $t_\mathrm{DSP}$), the astute reader will have realized that our population models contain either very few or no normal Pop~III stars.  Constructing a consistent picture of the chemical evolution of the Universe in these cases becomes somewhat more problematic than in the standard situation, where supernovae produced by the deaths of the original Pop~III stars provide the necessary chemical enrichment of gas to facilitate the formation of Pop~II stars.  Stars as heavy as the 800\,$M_\odot$ dark stars we consider here are typically expected to collapse directly to black holes \citep[e.g.][]{Umeda09}, producing very few metals. Because only stars in the mass range up to 260 $M_\odot$ produce metals \citep{Heger:02,Venkatesan:03b}, this constrains the mass range of a primordial stellar population that must necessarily seed the conditions for Pop~II star formation (and this is used to justify the Pop~III mass function we consider below). Moreover, the yields of heavy elements from supernovae from the first stars as well as the true distribution of metallicities in the very first Pop~II stars are currently not well-constrained, so it is conceivable that a small number of highly-efficient Pop~III stars could provide the bulk of chemical enrichment necessary to allow the transition to Pop~II (and such a Pop~II may have begun somewhat later in some locations, and with a somewhat smaller metallicity than is considered typical).  In any case, stars as massive as 800\,$M_\odot$ operate very close to the Eddington luminosity, so are expected to exhibit very strong stellar winds and experience numerous mass-loss events. Although mass loss from metal-poor or metal-free stars is expected to be substantially reduced in comparison to that from their metal-rich cousins, depending upon the rotational and convective properties of the first stars, the material blown off from such objects may well be sufficiently processed to also contribute to the chemical enrichment of the Universe. 

\subsection{Ionizing photon fluxes}

At each timestep of our reionization calculation, we calculate the weighted-average, mass-normalized, hydrogen-ionizing photon output of the combined stellar population as
\begin{equation}
Q_\mathrm{tot}(t) = f_\mathrm{DS}Q_\mathrm{DS}(t) + (1-f_\mathrm{DS})Q_\mathrm{normal}(t).
\end{equation}
Here $Q_\mathrm{DS}(t)$ refers to the ionizing photon output per unit mass of stars in the population originally consisting of dark stars.  Depending upon the time $t$ in question, and the values of the lifetime parameters $t_\mathrm{DSP}$ and $t_\mathrm{DSNMS}$, this may be equal to either $Q_\mathrm{DSP}$, $Q_\mathrm{DSNMS}$, $Q_\mathrm{Pop\,II}$ or $Q_\mathrm{Pop\,III}$.  $Q_\mathrm{normal}(t)$ refers to the population originally consisting of normal Pop III stars, and is equal to $Q_\mathrm{Pop\,III}$ for $t<10$\,Myr, and $Q_\mathrm{Pop\,II}$ for $t\ge10$\,Myr.

Explicitly, if we designate $t_0$ as the time elapsed since the onset of star formation (DS and/or Pop III),  and define the specific averaged $Q_\mathrm{H}$ factors
\begin{eqnarray}
\label{QA}
Q_\mathrm{tot,DSP+P3}\ =&\ (1-f_\mathrm{DS})Q_\mathrm{Pop\,III}\\
\label{QB}
Q_\mathrm{tot,DSP+P2}\ =&\ (1-f_\mathrm{DS})Q_\mathrm{Pop\,II}\\
\label{QC}
Q_\mathrm{tot,DSNMS+P3}\ =&\ f_\mathrm{DS}Q_\mathrm{DSNMS} + (1-f_\mathrm{DS})Q_\mathrm{Pop\,III}\\
\label{QD}
Q_\mathrm{tot,DSNMS+P2}\ =&\ f_\mathrm{DS}Q_\mathrm{DSNMS} + (1-f_\mathrm{DS})Q_\mathrm{Pop\,II},
\end{eqnarray}
we have three possible scenarios, depending upon the relative values of the lifetime parameters $t_\mathrm{DSP}$, $t_\mathrm{DSNMS}$ and $t_\mathrm{Pop\,III}$:
\begin{itemize}
\item $t_\mathrm{DSP} + t_\mathrm{DSNMS} < t_\mathrm{Pop\,III}$ 
  \begin{enumerate}
  \item $Q_\mathrm{H} = Q_\mathrm{tot,DSP+P3}$, duration $t_\mathrm{DSP}$ 
  \item $Q_\mathrm{H} = Q_\mathrm{tot,DSNMS+P3}$, duration $t_\mathrm{DSNMS}$ 
  \item $Q_\mathrm{H} = Q_\mathrm{Pop\,III}$, duration $t_\mathrm{Pop\,III} - (t_\mathrm{DSNMS} + t_\mathrm{DSP})$ 
  \item $Q_\mathrm{H} = Q_\mathrm{Pop\,II}$, duration $t_0 - t_\mathrm{Pop\,III}$ 
  \end{enumerate}
\item $t_\mathrm{DSP} < t_\mathrm{Pop\,III}$ and $t_\mathrm{DSP} + t_\mathrm{DSNMS} > t_\mathrm{Pop\,III}$
  \begin{enumerate}
  \item $Q_\mathrm{H} = Q_\mathrm{tot,DSP+P3}$, duration $t_\mathrm{DSP}$ 
  \item $Q_\mathrm{H} = Q_\mathrm{tot,DSNMS+P3}$, duration $t_\mathrm{Pop\,III} - t_\mathrm{DSP}$ 
  \item $Q_\mathrm{H} = Q_\mathrm{tot,DSNMS+P2}$, duration $t_\mathrm{DSNMS} - (t_\mathrm{Pop\,III} - t_\mathrm{DSP})$ 
  \item $Q_\mathrm{H} = Q_\mathrm{Pop\,II}$, duration $t_0 - (t_\mathrm{DSP} + t_\mathrm{DSNMS})$ 
  \end{enumerate}
\item $t_\mathrm{DSP} > t_\mathrm{Pop\,III}$
  \begin{enumerate}
  \item $Q_\mathrm{H} = Q_\mathrm{tot,DSP+P3}$, duration $t_\mathrm{Pop\,III}$ 
  \item $Q_\mathrm{H} = Q_\mathrm{tot,DSP+P2}$, duration $t_\mathrm{DSP} - t_\mathrm{Pop\,III}$ 
  \item $Q_\mathrm{H} = Q_\mathrm{tot,DSNMS+P2}$, duration $t_\mathrm{DSNMS}$ 
  \item $Q_\mathrm{H} = Q_\mathrm{Pop\,II}$, duration $t_0 - (t_\mathrm{DSP} + t_\mathrm{DSNMS})$ 
  \end{enumerate}
\end{itemize}
For all scenarios, and in Eqs.~\ref{QA} and \ref{QB} above, we assume $Q_\mathrm{DSP}=0$ during the DSP phase.  

\begin{deluxetable}{l l c r r}
\centering
\tablecaption{Ionizing photon fluxes, with $Q_\mathrm{H}$ values high enough to contribute to reionization \citep[cf.~Table 1 in][]{Tumlinson:04} marked in bold.
\label{QHtable}}
\tablewidth{0pt}
\tablehead{
\colhead{DM Mass} & \colhead{Age} & \colhead{Stellar} & \colhead{$Q_{\rm H}$} & \colhead{$Q_{\rm H}$}\\ 
\colhead{} & \colhead{(yr)} & \colhead{Mass (M$_\odot$)} & \colhead{(s$^{-1}$)} & \colhead{(s$^{-1}$ M$_\odot^{-1}$)}} 
\startdata
1 GeV & $3.1\times10^5$ & 756\phantom{$^\dagger$} & 1.05$\times10^{44}$ & 1.38$\times10^{41}$ \\
1 GeV & $3.3\times10^5$ & 793\phantom{$^\dagger$} & {\bf 4.26$\times10^{49}$} & {\bf 5.37$\times10^{46}$} \\
1 GeV & $4.6\times10^5$ & 824\phantom{$^\dagger$} & {\bf 5.37$\times10^{50}$} & {\bf 6.52$\times10^{47}$} \\
 &   & & \\
100 GeV & $3.0\times10^5$ & 716\phantom{$^\dagger$}  & 6.28$\times10^{47}$ & 8.78$\times10^{44}$ \\
100 GeV & $3.9\times10^5$ & 779$^\dagger$ & {\bf 4.97$\times10^{50}$} & {\bf 6.38$\times10^{47}$}\\
100 GeV & $4.1\times10^5$ & 787$^\star$  & {\bf 5.20$\times10^{50}$} & {\bf 6.61$\times10^{47}$} \\
 &   & & \\
$10$ TeV & $0.9\times10^5$  & 327\phantom{$^\dagger$} & 3.40$\times10^{43}$ & 1.04$\times10^{41}$ \\
$10$ TeV & $2.7\times10^5$ & 553\phantom{$^\dagger$} & {\bf 3.27$\times10^{50}$} & {\bf 5.92$\times10^{47}$} \\
 &   & & \\
\multicolumn{2}{l}{Normal Pop~II} & 1--100\phantom{$^\dagger$} &  & {\bf 7.76 $\times10^{46}$} \\
\multicolumn{2}{l}{Normal Pop~III} & 10--140\phantom{$^\dagger$} &  & {\bf 4.30 $\times10^{47}$} 
\enddata
\tablenotetext{$\dagger$}{This case corresponds to the NC scenario considered in this work, and does not include any capture by nuclear scattering.  All other entries in this table, including the model used for the MC and EC scenarios, correspond to models where nuclear scattering provides a similar amount of power to DM obtained by gravitational contraction \citep[see][for details]{Spolyar09}.}
\tablenotetext{$\star$}{This case is used in calculations for the EC and MC scenarios considered in this work.}
\end{deluxetable}

We calculate ionizing photon fluxes during the DSNMS phase using model dark star atmospheres computed with TLUSTY \citep{TLUSTY}, as described in \citet{Zackrisson10a}.  Hydrogen-ionizing photon fluxes for some example dark star models computed by \citet{Spolyar09} are shown in Table \ref{QHtable}.  Here we display $Q_\mathrm{H}$ values for three different stellar models, computed assuming three different DM masses.  We show snapshots of $Q_\mathrm{H}$ at different times in the respective models' evolution: larger stellar masses correspond to later models, as dark stars gradually accrete more matter. We do not show $Q_\mathrm{H}$ values for earlier times in the simulations (corresponding to the main part of the DSP phase), as the ionizing fluxes of the earliest (lowest-mass) snapshots shown in Table \ref{QHtable} are already too low to be significant for reionization.  As is to be expected, ionizing photon fluxes increase with time and stellar mass as the dark stars become hotter, more compact and luminous as they move from DSP to DSNMS, and finally, to the zero-age main sequence.  

For a fixed initial DM density, larger DM masses lead to slightly decreased energy production in the stellar core (due to the decreased number density implied by a constant mass density).  This can be seen in the lower final mass of the $m_\chi=1$\,TeV model in Table \ref{QHtable}, where DM annihilation has extended the accretion phase during the DSP less than it would if the DM mass were smaller \citep[see][for more details]{Spolyar09}, resulting in a reduced $Q_\mathrm{H}$ during the DSNMS phase.  In general however, the mass of the DM particle has only a weak impact upon the phenomenology of dark stars.  For the remainder of this paper, we focus on the example case $m_\chi = 100$\,GeV.

For the NC scenario, we use the final, 779\,$M_\odot$, $m_\chi = 100$\,GeV stellar model of \citet{Spolyar09}, which was computed without including any capture of WIMPs via nuclear scattering.  When calculating ionizing fluxes during the DSNMS phase in the MC and EC scenarios, we instead use the corresponding 787\,$M_\odot$, $m_\chi=100$\,GeV model, which included a small amount of DM capture by nuclear scattering (see \citeauthor{Spolyar09}~for details).  In practice, there is very little difference between the models of \citet{Spolyar09} with and without the very small amount of capture they included.  

The ionizing photon fluxes of our canonical Pop~II and III populations are also given in Table \ref{QHtable}.

\section{Reionization Calculations}
\label{reionization}

We use the semianalytic reionization model in \citet{Venkatesan:03} for a $\Lambda$CDM cosmology. The growth of ionized regions is tracked by a Press-Schechter
formalism in combination with numerical solutions for the
growth of individual ionization fronts.  We take our cosmological parameter set from the latest WMAP7 results \citep{Larson:10}.\footnote{We used the WMAP7 cosmological parameter set at: \protect\href{http://lambda.gsfc.nasa.gov/product/map/current/params/lcdm_sz_lens_wmap7.cfm}{http://lambda.gs-fc.nasa.gov/product/map/current/params/lcdm\_sz\_lens\_wmap7.cfm}}
We assume that the fraction of baryons forming stars in each halo is $f_\star = 0.05$, and that the escape fraction of H~II ionizing radiation from halos is $f_{\rm esc} = 0.1$. Theoretical calculations, both analytical and from numerical simulations, as well as observations of star-forming galaxies in the local and high-$z$ Universe, indicate that $f_\star = 0.01-0.1$, and $f_{\rm esc} = 0.01-0.2$ (\citealt{Venkatesan:03}, and references therein). 

We allow star formation in all halos of virial temperature \mbox{$\ga$ 375~K} (rather than $10^4$~K as in \citealt{Venkatesan:03}, but similar to $10^3$~K in \citealt{Tumlinson:04}) starting at $z=20$. This ensures that in our cases where $f_{\rm DS} = 1$, there are  sufficient baryons in a 10$^6$ M$_\odot$ dark matter halo at $z \sim$ 20 for our
adopted star formation efficiencies to form a single 800 M$_\odot$ dark star. This relies on the assumption that halos with virial temperatures $\la 10^4$~K can cool effectively using, e.g., H$_2$ or other pathways \citep{Abel:02, Haiman:03}. \footnote{\label{note8}The assumption that each new halo forms metal-free stars breaks down at \mbox{$z \sim$ 8--9}, when the earliest small halos, presumably metal-enriched from their first episodes of Pop III star formation, begin to coalesce into larger objects.  Most of the scenarios considered here reionize earlier than this; those that do not would arguably not follow the standard timeline of chemical enrichment anyway, as the bulk of baryons would be tied up in long-lived DSP-phase dark stars at redshifts $z\ga9$, leading to some delay in the chemical evolution of the Universe.}  Increasing the virial temperature threshold (i.e. the minimum mass scale of collapsing halos) to values higher than our choice here will delay reionization correspondingly.  As the most interesting constraints we produce are based on dark-star induced delays to reionization, this choice is a conservative one.

We assume Pop III (metal-free) stars can form starting at $z=$20, in a Salpeter-slope initial mass function (IMF) spanning the range 10--140 M$_\odot$. This Pop III IMF\footnote{The Pop III IMF is independent of the presence of dark stars in our formalism, and is not designed to reflect the late stages of dark star evolution in any way.  As noted earlier, dark stars die at the end of the DSNMS phase, which is effectively their main sequence.} is one lacking in low-mass stars, and was motivated by ionization constraints and observations of abundance trends in metal-poor Galactic halo stars \citep{Tumlinson:04}.  It has a mass-normalized ionizing photon flux about a factor of 3 smaller than the ionizing flux of an IMF containing very massive stars (10$^2$--10$^3$ M$_\odot$, \citealt{Bromm:01}). We assume that the Pop III phase lasts for 10\,Myr in each halo, in agreement  with the duration of metal-free star formation calculated from numerical simulations of halo self-enrichment \citep{Wada:03}, and of interhalo enrichment \citep{Bromm:03, Tumlinson:04}. Subsequent to 10\,Myr after the onset of star formation in each halo,
the ionizing spectrum is switched to a representative example
of Pop II stars in a Salpeter-slope IMF in the mass range 1--100  M$_\odot$ with metallicity $Z =$ 0.001 \citep{Leitherer:99}. 

It is possible that Pop~II star formation need not always follow that of Pop~III; both populations could form simultaneously in separate gas clouds within individual halos, due to inhomogeneous metal enrichment. However, this is most likely to happen in more massive halos, as smaller halos will experience rapid self-enrichment or lose their gas entirely after the first Pop~III supernovae. Massive halos are also more likely to be made up of smaller halos with disparate chemical evolution histories (see footnote \ref{note8}). Such massive halos are more common at $z \sim$ 7--9, when the Universe is 0.5\,Gyr old. This cosmic age is substantially  larger than the 10\,Myr self-enrichment timescales of halos discussed above, over which the conditions to form Pop III are lost. Additionally, the gain in hydrogen-ionizing photon production from Pop III relative to Pop II is a factor of order 0.6 up to a few \citep{Tumlinson:04}. When one considers that this could occur only in larger halos for at most 10\,Myr (at a cosmic age of 0.5--1\,Gyr), the impact on reionization should be relatively small. We have confirmed in separate calculations for such a scenario that this is the case. We therefore do not consider simultaneous Pop III and Pop II star formation within the same halo in our models.

Note that in our formalism for using weighted-average ionizing photon fluxes from Pop~III, Pop~II and dark stars, at each cosmological epoch, we count the time-appropriate $Q_\mathrm{H}$ at that epoch arising from all star-forming halos that have collapsed over a range of redshifts starting at $z=20$.  Reionization is defined as the overlap of individual
ionized regions of H\,\textsc{ii}, when its  IGM volume filling factor equals unity \citep{Venkatesan:03}.

\begin{figure*}[tbp]
\includegraphics[width=0.95\columnwidth, trim = 0 20 0 0, clip=true]{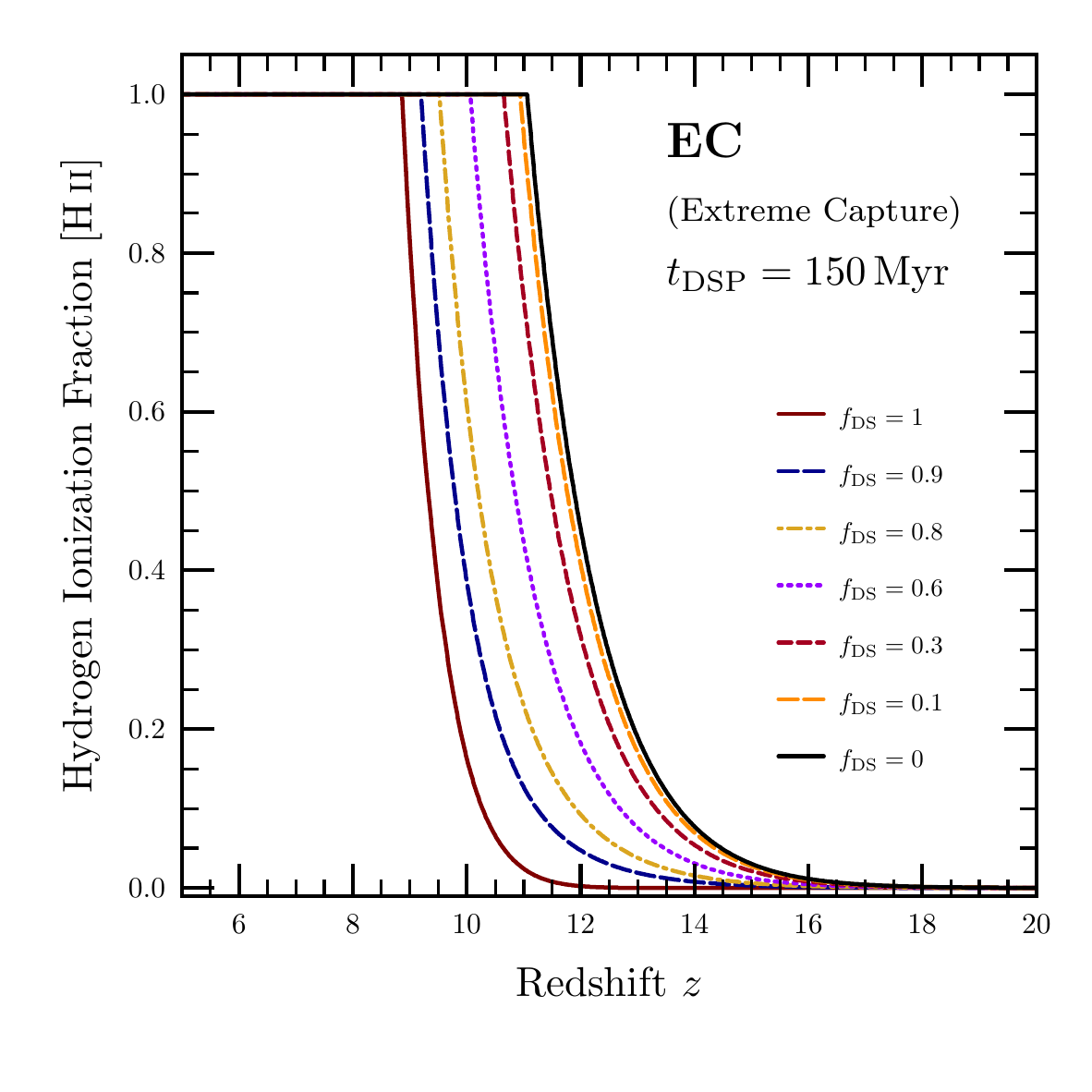}\hspace{8mm}
\includegraphics[width=0.95\columnwidth, trim = 0 20 0 0, clip=true]{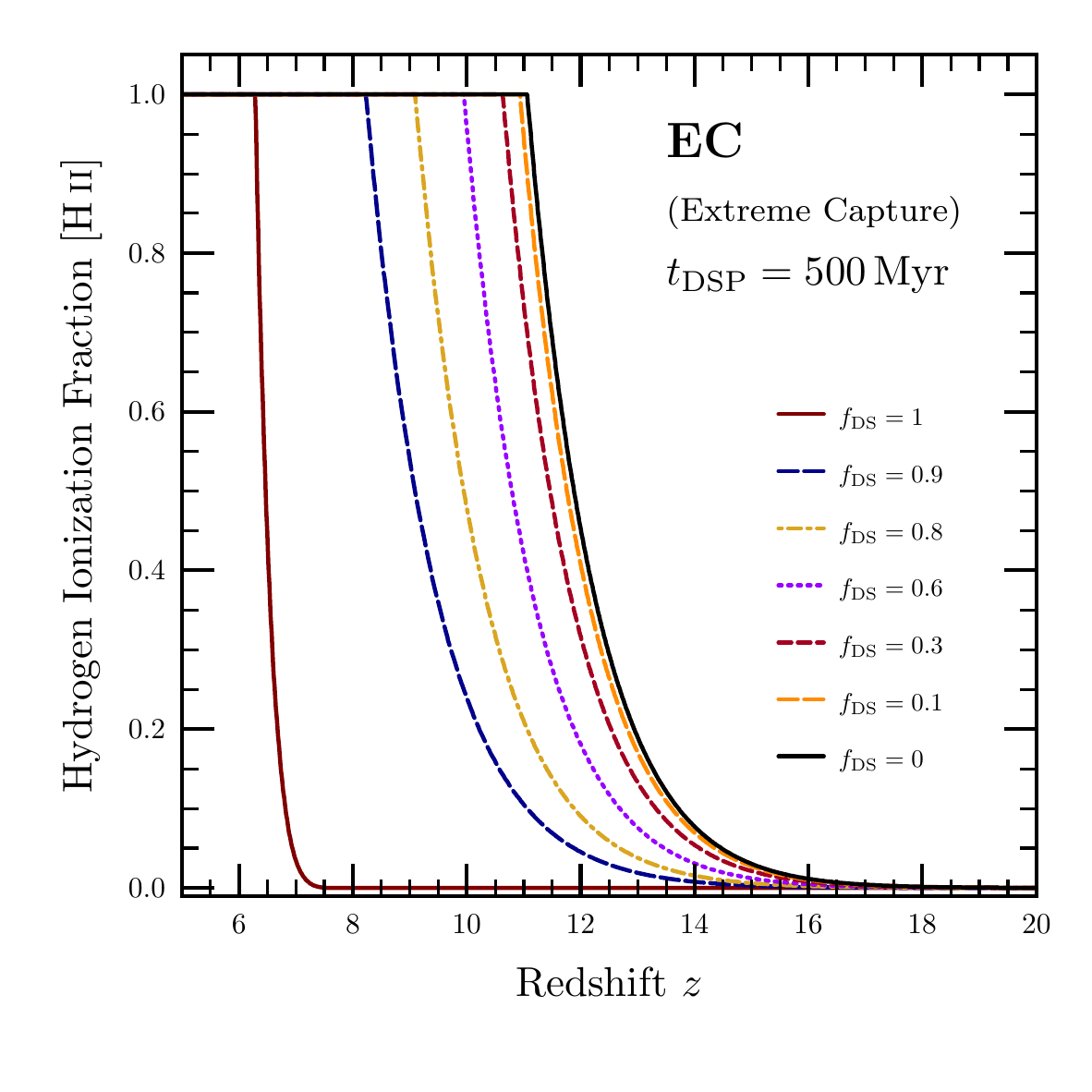}\\
\includegraphics[width=0.95\columnwidth, trim = 0 20 0 0, clip=true]{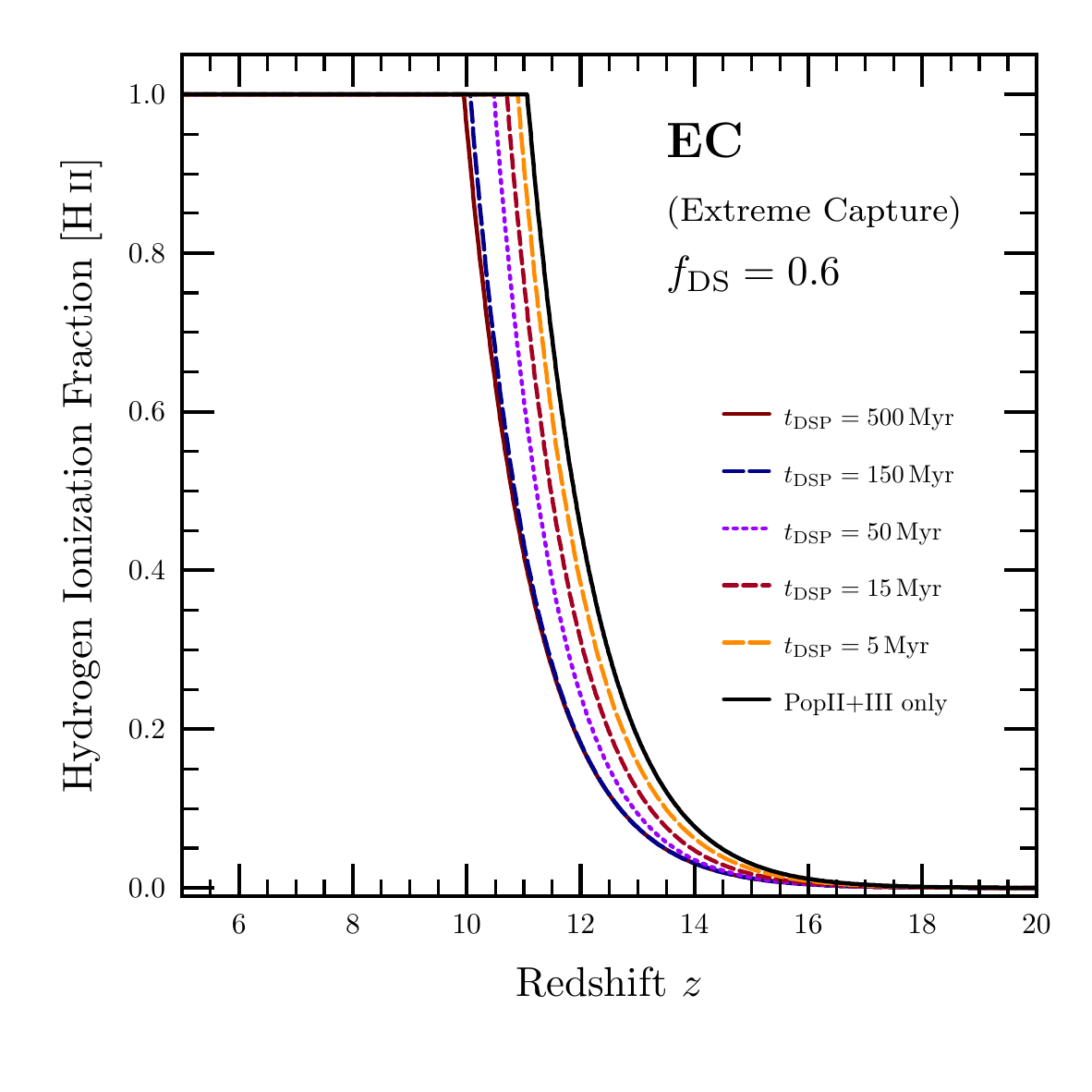}\hspace{8mm}
\includegraphics[width=0.95\columnwidth, trim = 0 20 0 0, clip=true]{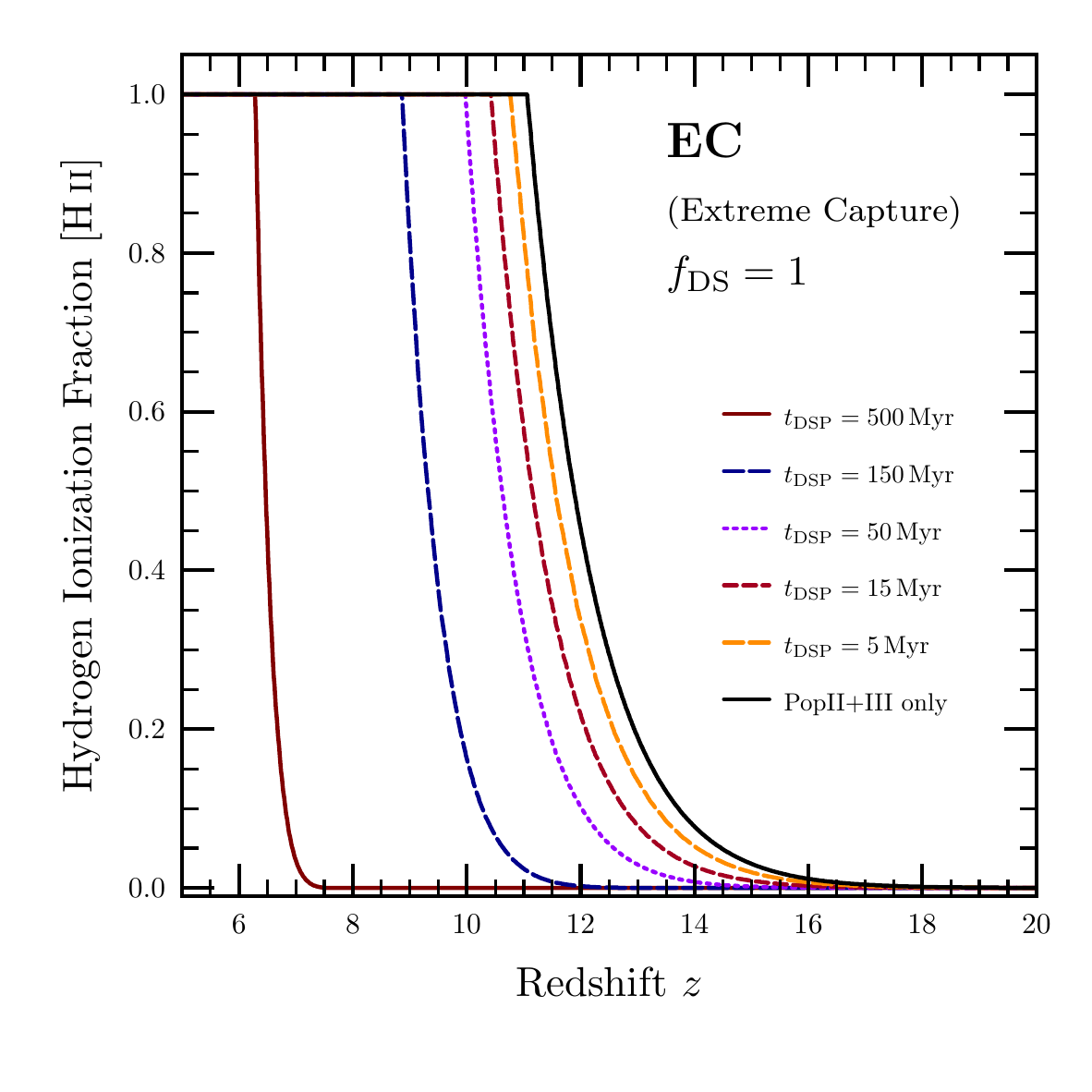}
\caption{Ionization histories of a Universe containing dark or semi-dark stellar populations, in the extreme capture (EC) scenario, where dark stars live extended lives as cool, diffuse objects.  Here we plot the fraction of hydrogen in H\,\textsc{ii} as a function of redshift.  Top panels compare histories for different dark star fractions $f_\mathrm{DS}$, in situations where dark stars live a long time in the DSP phase ($t_\mathrm{DSP} = 150$\,Myr and $500$\,Myr).  Bottom panels compare histories for different DSP lifetimes, in situations where dark stars make up a substantial fraction of the mass budget of the first population of stars ($f_\mathrm{DS}=0.6$ and $1$).  Longer-lived and more numerous EC dark stars delay reionization.  Here we have assumed that the fraction of the initial star-forming baryonic mass budget of the Universe that \textit{does not} form dark stars instead forms normal Pop~III stars, which die after 10\,Myr and are replaced by newly-born Pop~II stars.  Following the death of the population of dark stars, and depending upon their time of death, they are replaced either directly with newborn Pop~II stars, or with newborn Pop~III stars, which themselves later die and get replaced by newborn Pop~II stars.}
\label{fig:EC_ion}
\end{figure*}

\begin{figure*}[tbp]
\includegraphics[width=0.95\columnwidth, trim = 0 20 0 0, clip=true]{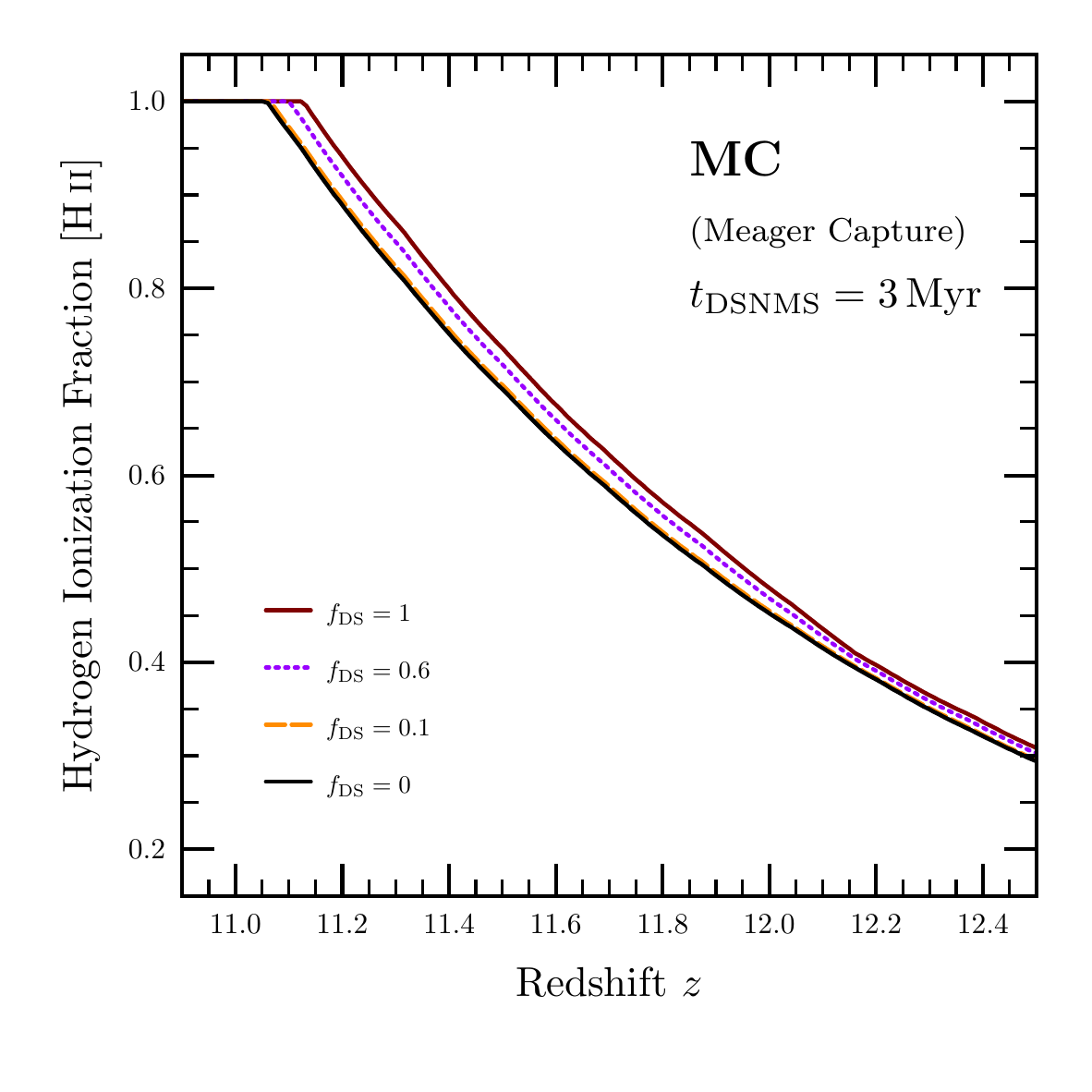}\hspace{8mm}
\includegraphics[width=0.95\columnwidth, trim = 0 20 0 0, clip=true]{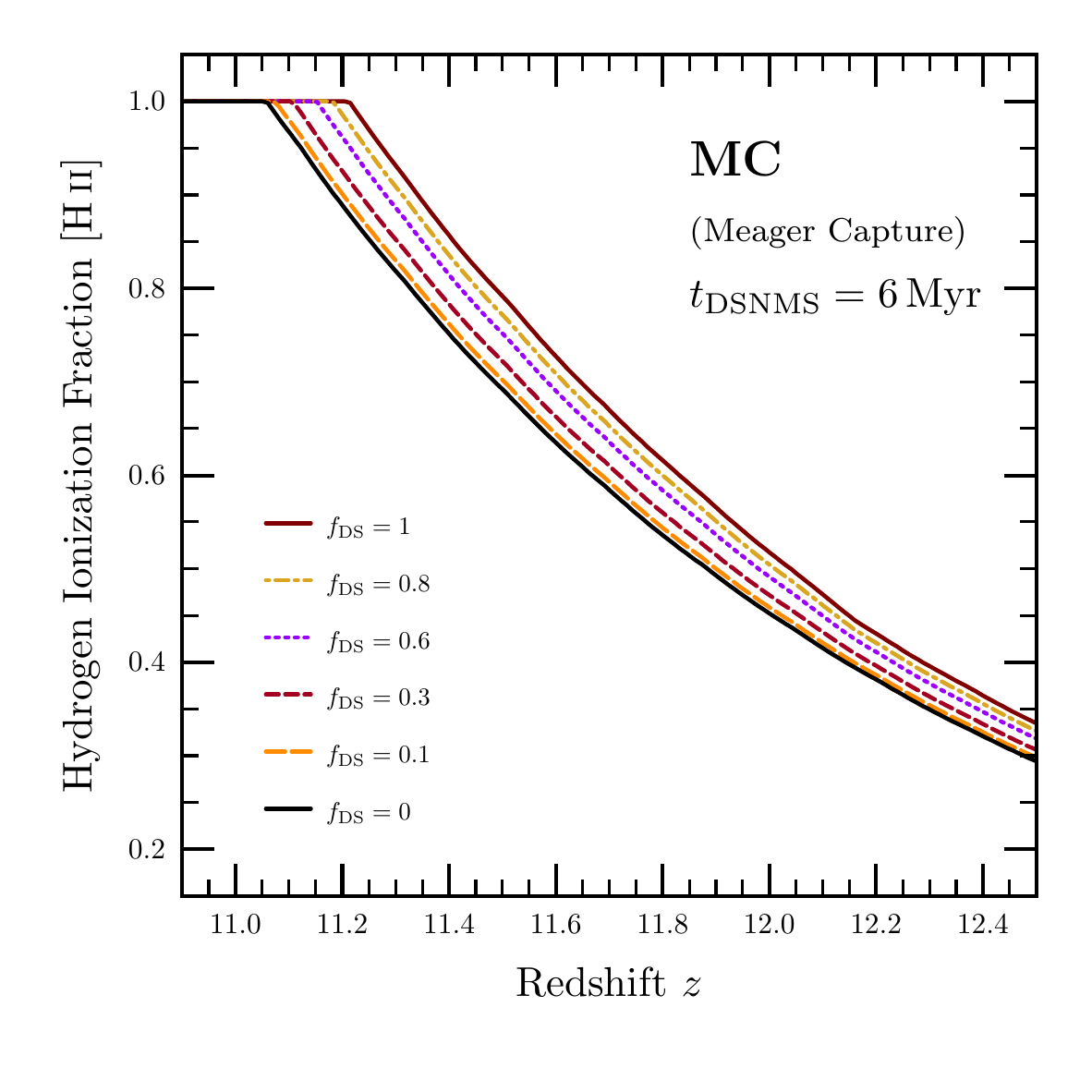}
\caption{Ionization histories of a Universe containing dark stars, in the meager capture (MC) scenario, where dark stars effectively receive a small extension to their main-sequence lifetimes.  The two panels compare histories for different dark star fractions $f_\mathrm{DS}$, where dark stars live in the DSNMS phase for close to the maximum time allowed by core hydrogen depletion ($t_\mathrm{DSNMS} = 3$ and $6$\,Myr).  Longer DSNMS lifetimes and larger dark star fractions lead to small increases in the speed of reionization (note the zoomed axes relative to Fig.~\protect\ref{fig:EC_ion}).  Non-dark aspects of the calculations are as described for Fig.~\protect\ref{fig:EC_ion}.}
\label{fig:MC_ion}
\end{figure*}

\begin{figure*}[tbp]
\includegraphics[width=0.95\columnwidth, trim = 0 20 0 0, clip=true]{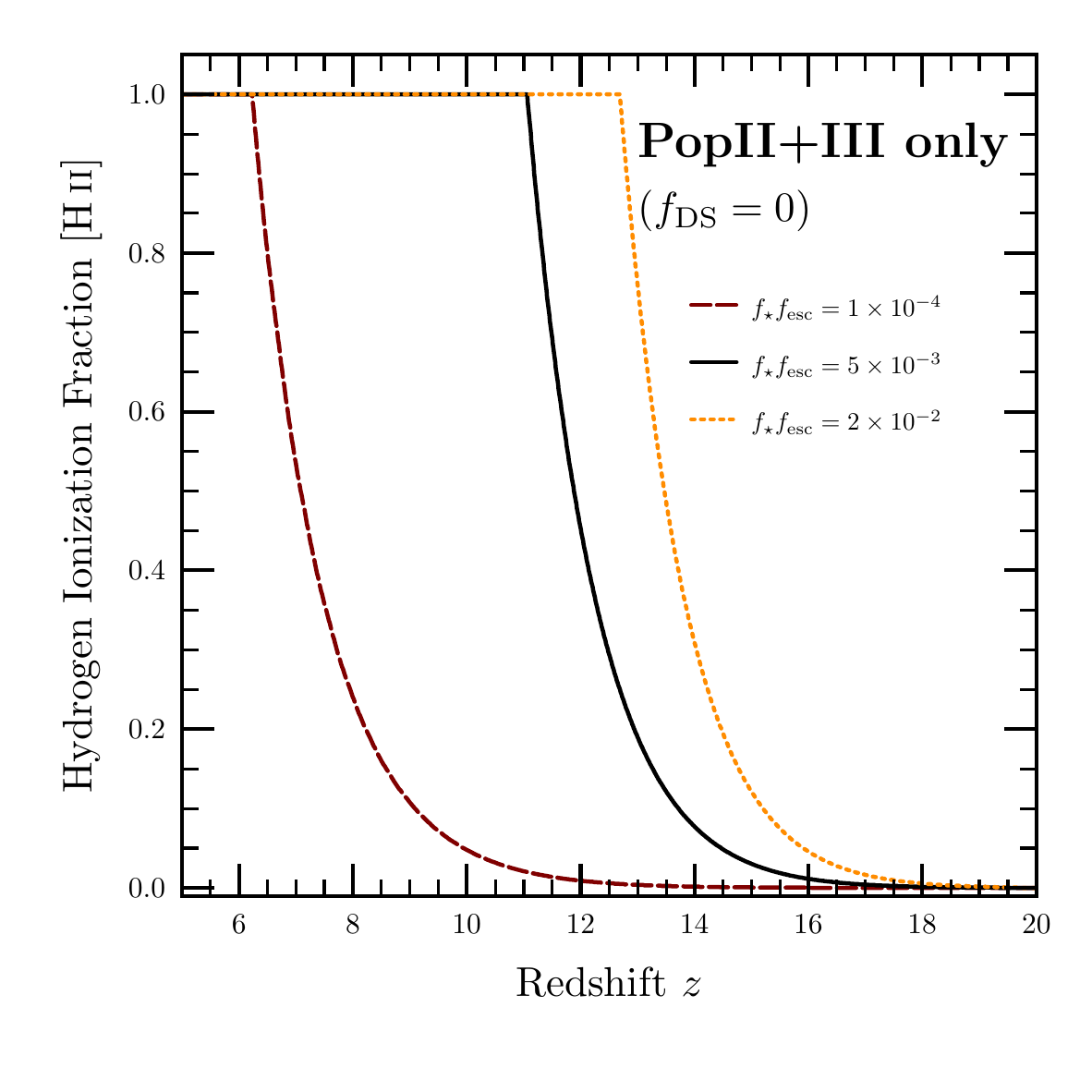}\hspace{8mm}
\includegraphics[width=0.95\columnwidth, trim = 0 20 0 0, clip=true]{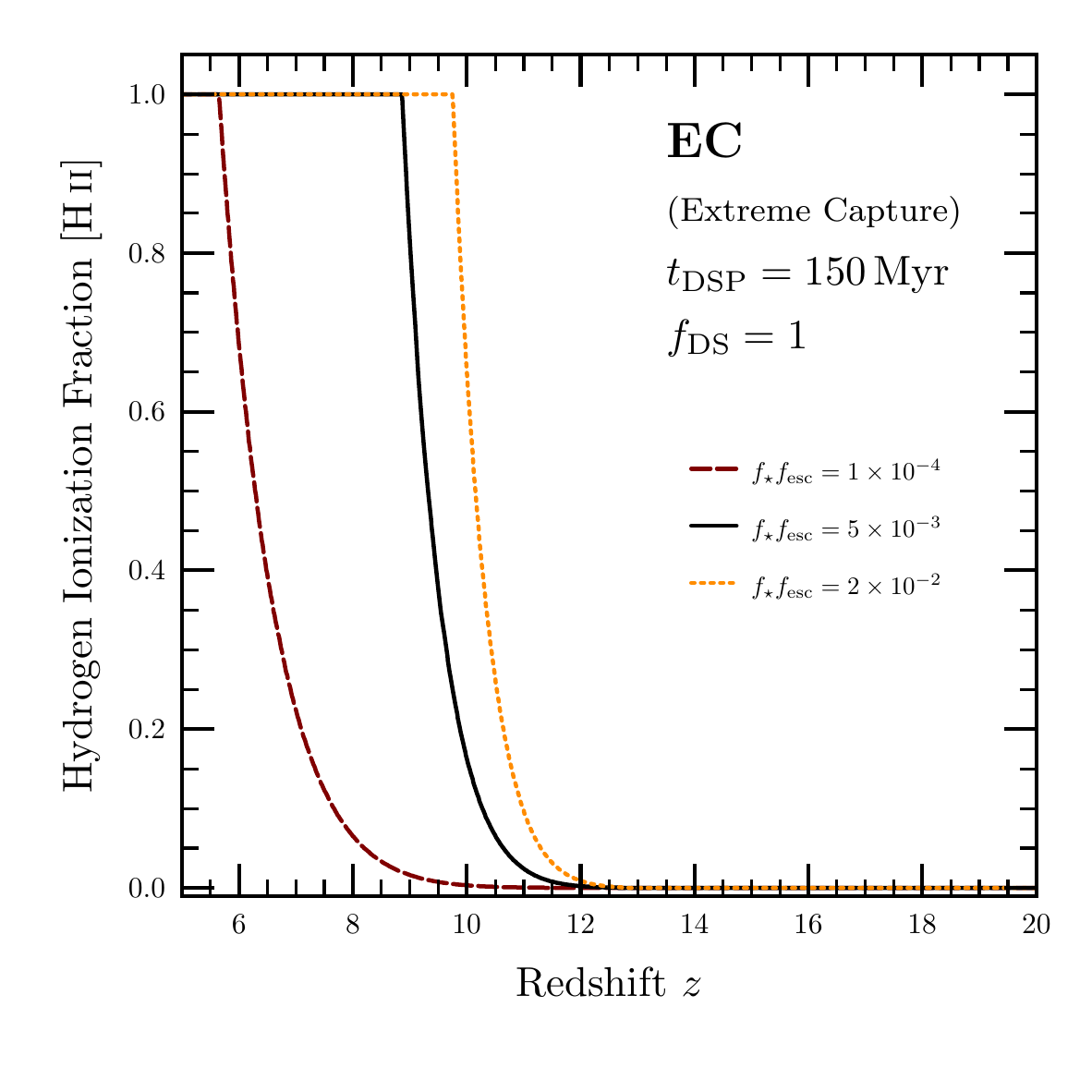}
\caption{Impacts of varying astrophysical parameters upon the reionization history of a Universe containing no dark stars (left), and one containing EC dark stars with $t_\mathrm{DSP} = 150$\,Myr, $f_\mathrm{DS}=1$ (right).  Here we show the effects of varying the product of the star-forming baryon fraction $f_\star$ and the ionizing photon escape fraction $f_\mathrm{esc}$.  The variation of astrophysical parameters induces a similar change in the reionization history of the Universe to dark stars (left), but has a slightly reduced impact when applied to reionization scenarios that include dark stars (right).  Variations in $f_\star f_\mathrm{esc}$ can not only delay reionization as EC dark stars do, but also speed it up, to a much greater extent than MC dark stars are able to do.}
\label{fig:AST_ion}
\end{figure*}

\section{Reionization history of a Universe containing dark stars}
\label{reionhist}

In the NC scenario, where dark stars are fueled by gravitationally-contracted dark matter only, we find that the reionization history of the Universe is effectively identical for all values of the dark star fraction $f_\mathrm{DS}$ (Table \ref{bigtable}).  This is despite the fact that during the first 0.4\,Myr of star formation, the dark stellar population contributes nothing to reionization, reducing $Q_\mathrm{tot}$ to \mbox{$(1-f_\mathrm{DS})Q_\mathrm{Pop\,III}$}.  In the ensuing DSNMS phase, the earlier lack of ionizing photons is compensated for by an excess relative to the canonical situation (where $f_\mathrm{DS}=0$ and there are no dark stars, only the normal Pop~III followed by Pop~II), because $Q_\mathrm{DSNMS} > Q_\mathrm{Pop\,III}$.  Although the net impacts of these two effects on the redshift of reionization cancel, one might expect a difference in the time-evolution of the ionization fraction relative to the simple Pop~III+II case; due to the short durations of the DSP and DSNMS phases in the NC scenario, this effect is however too small to notice.

In Fig.~\ref{fig:EC_ion}, we show ionization histories in the extreme capture (EC) scenario, for various combinations of $f_\mathrm{DS}$ and the DSP lifetime $t_\mathrm{DSP}$.  Here we see a marked effect on reionization, with populations containing large numbers of dark stars and/or relatively long-lived ones resulting in substantially delayed reionization.  As expected, the larger the values of $f_\mathrm{DS}$ and $t_\mathrm{DSP}$, the larger the effect.   For larger $f_\mathrm{DS}$ and $t_\mathrm{DSP}$, $z_{\rm reion}$ occurs later, as $Q_\mathrm{tot}$ remains at lower values for longer periods of time during the earliest part of the star-formation history in each halo.  For longer delays, reionization occurs more quickly as IGM ionization becomes able to build up more quickly. This is a result of two effects: the main sources of ionizing photons turning on at lower redshifts,  and the fact that the IGM density decreases rapidly with decreasing redshift, leading to increasing IGM recombination timescales at later cosmic times. This explains why the rapidity of reionization becomes more pronounced with increasing $t_\mathrm{DSP}$ than with increasing $f_\mathrm{DS}$.

Comparing our results in Table~\ref{bigtable} with the constraints from WMAP7, we find that a few of our cases can be immediately ruled out by simply having too low a value of $\tau_\mathrm{e}$, the optical depth to electron scattering, or $z_{\rm reion}$. The global fit to WMAP7 and other cosmological data \citep{WMAP7} implies that the integrated optical depth back to the surface of last scattering (i.e.~recombination at $z \sim 1090$), is $\tau_ \mathrm{e} \sim 0.088 \pm 0.014$.  For a simple step function reionization history, this is $z_{\rm reion} = 10.6 \pm 1.2$. A direct comparison with Table~\ref{bigtable} would in principle rule out all cases with $\tau_\mathrm{e} \la 0.074$, or with $z_{\rm reion} \la 9.4$. Note that the $\tau_\mathrm{e}$ constraint is more direct, and that limits based on $z_{\rm reion}$ are only approximate owing to the WMAP7 assumption of a simplified reionization history (step function ionization at fixed $z_{\rm reion}$). Our models are more realistic and track the details of reionization with variations in galaxy halo masses and astrophysical parameters, so that our derived $z_{\rm reion}$ and that from WMAP7 may not be directly comparable. Thus, comparing the values of $\tau_\mathrm{e}$ from Table~\ref{bigtable} to the WMAP7 limits, we see that the EC case with $t_{\rm DSP} = 500$\,Myr and $f_{\rm DS} = 1$ is ruled out. More EC cases, e.g. $t_{\rm DSP} = 150$\,Myr and $f_{\rm DS} = 0.9-1$, or $t_{\rm DSP} = 500$\,Myr and $f_{\rm DS} \ga 0.75$, would be ruled out at face value if we were to compare directly with the WMAP7 limit that $z_{\rm reion} = 9.4-11.8$.  Similarly, one might also conclude that many models with low $t_{\rm DSP}$ and $f_{\rm DS}$ are ruled out for producing $z_{\rm reion}$ and $\tau_\mathrm{e}$ \textit{exceeding} the upper limit of the WMAP7 error band.  Whilst this is indeed true when $f_\star f_\mathrm{esc}=0.005$ as we assume here, such limits are not really robust to variations in astrophysical parameters.  We discuss integrated optical depths and corresponding constraints in more detail in the following section.

For the meager capture (MC) scenario, the effects are less dramatic; in Fig.~\ref{fig:MC_ion} we show a zoomed-in section of the full history of reionization, for a few combinations of $f_\mathrm{DS}$ and the DSNMS lifetime $t_\mathrm{DSNMS}$.  Here, increasing $f_\mathrm{DS}$ and $t_\mathrm{DSNMS}$ results in progressively earlier reionization, the exact opposite trend relative to the EC cases.  This is because $Q_\mathrm{DSNMS} > Q_\mathrm{Pop\,III}$, so extending the DSNMS phase and increasing the fraction of baryons contained in it causes reionization to happen more quickly than with only a normal Pop~III IMF.  Again, as expected the effects increase with increasing $f_\mathrm{DS}$ and $t_\mathrm{DSNMS}$.  The reason the MC scenario has a much smaller effect on reionization than the EC scenario is that its duration is already much more strongly constrained, in this case by the fusion-burning timescale of core hydrogen during the DSNMS phase.

In Fig.~\ref{fig:AST_ion} we show the impact of varying the product of the astrophysical parameters $f_\star$ and $f_\mathrm{esc}$ from our canonical value of $f_\star f_\mathrm{esc}$ = 0.005 to the extreme values of 0.02 ($f_\star$ = 0.1, $f_\mathrm{esc}$ = 0.2) or $10^{-4}$ ($f_\star$ = $f_\mathrm{esc}$ = 0.01).  We give the resulting reionization histories both for a standard Pop~III without dark stars, and for an EC example  with $f_\mathrm{DS}=1, t_\mathrm{DSP}=150$\,Myr.  Here, we see that within this range of astrophysical uncertainties, a large range of reionization histories is possible.  Indeed, in the most extreme cases, dark stars have a similar magnitude effect as the variation of astrophysical parameters.  This degeneracy is unfortunate, but not unexpected; substantial uncertainty exists in reionization models at present, even before introducing the possibility of stellar populations including dark stars.  Hearteningly however, the impact of the astrophysical uncertainties is reduced in situations where dark stars play a significant role, as can be seen by comparing the two panels of Fig.~\ref{fig:AST_ion}.

The results we present here agree broadly with those of \citet{Schleicher08}, but the correspondence is not immediately obvious.  \citeauthor{Schleicher08}~considered reionization from `MS-dominated' (main sequence), and `CD' (capture dominated) dark stars, roughly corresponding to our own MC and EC scenarios, respectively.  They investigated the case where $f_\mathrm{DS}=1$, showing that the higher ionizing photon fluxes of the DSNMS phase hasten reionization, whereas the lower fluxes of the DSP phase delay it.  This is in good agreement with what we show here, and earlier predictions by \citet{Yoon08}.  Where we differ from \citeauthor{Schleicher08}'s analysis is in our atmospheric modeling (we use actual model atmospheres rather than black-body spectra), and in the details of our population modeling.

Here we carefully treat the allowed lifetimes of the different dark star phases, taking into account existing limits from the timescales of core hydrogen burning and self-annihilation of the dark matter halos surrounding dark stars.  \citet{Schleicher08} adopted the ionizing photon fluxes of \citet{Yoon08}, which accounted for the limit from core hydrogen burning, but not from halo self-annihilation or disruption (although both did at least acknowledge that this might be a concern).  Many of these lifetimes we now know to be theoretically inaccessible due to self-annihilation constraints (see the paragraph discussing viable ranges of $t_\mathrm{DSP}$ and $t_\mathrm{DSNMS}$ in Sec.~\ref{popmodels}).  Many of the models \citeauthor{Schleicher08} considered were accordingly ruled out by the redshifts of reionization or integrated optical depths.  \citet{Schleicher08} also allowed only one or the other of the DSP or DSNMS phases in their calculations, whereas we include both self-consistently.

\citet{Schleicher08} went on to investigate whether the effects of dark stars can be compensated for by more complicated reionization histories, varying assorted astrophysical parameters and introducing a second period of recombination followed by a late starburst, leading to a two-stage reionization history.  Here we have shown that although dark stars \textit{can} have a significant impact upon the reionization history of the Universe, they need not \textit{necessarily}.  Even in cases with the most extreme effects (e.g.~EC scenarios with $f_\mathrm{DS}=1$, $t_\mathrm{DSP}>100$\,Myr), Fig.~\ref{fig:AST_ion} shows that ad hoc scenarios like those considered by \citeauthor{Schleicher08}~are not necessary to reconcile dark stars with reionization constraints; a simple increase in $f_\star f_\mathrm{esc}$ does the job quite well enough.

\begin{figure*}[tbp]
\includegraphics[width=0.95\columnwidth, trim = 0 20 0 0, clip=true]{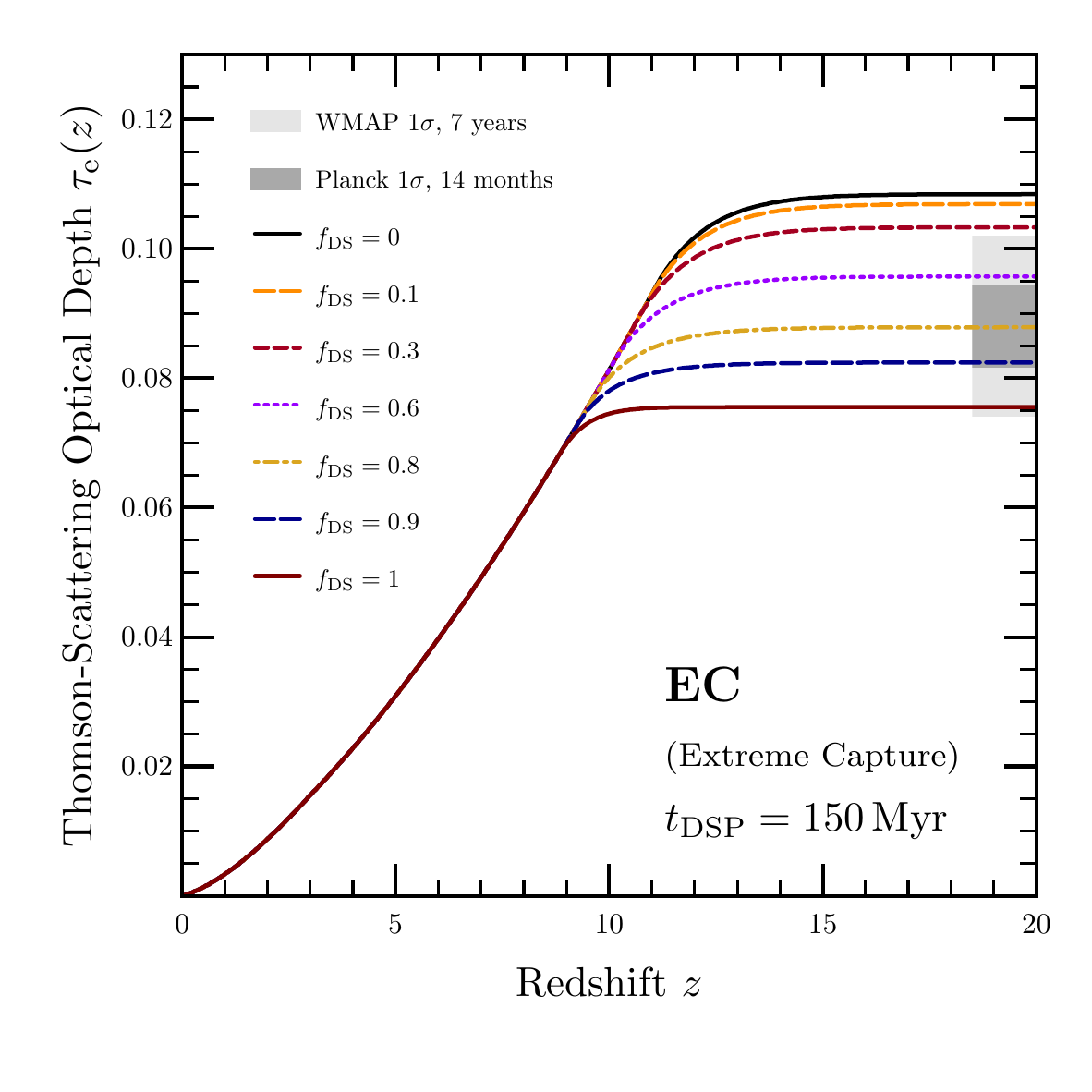}\hspace{8mm}
\includegraphics[width=0.95\columnwidth, trim = 0 20 0 0, clip=true]{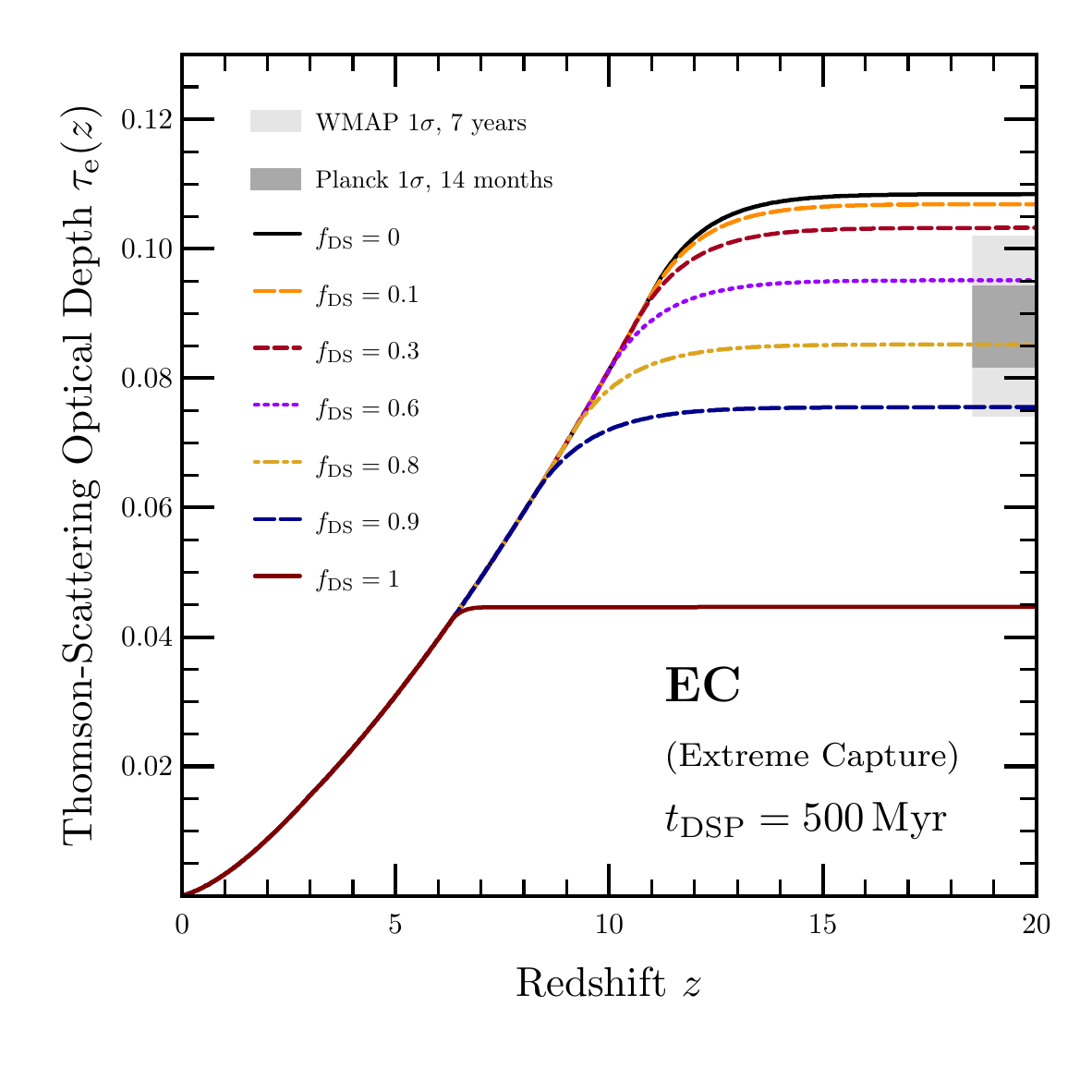}\\
\includegraphics[width=0.95\columnwidth, trim = 0 20 0 0, clip=true]{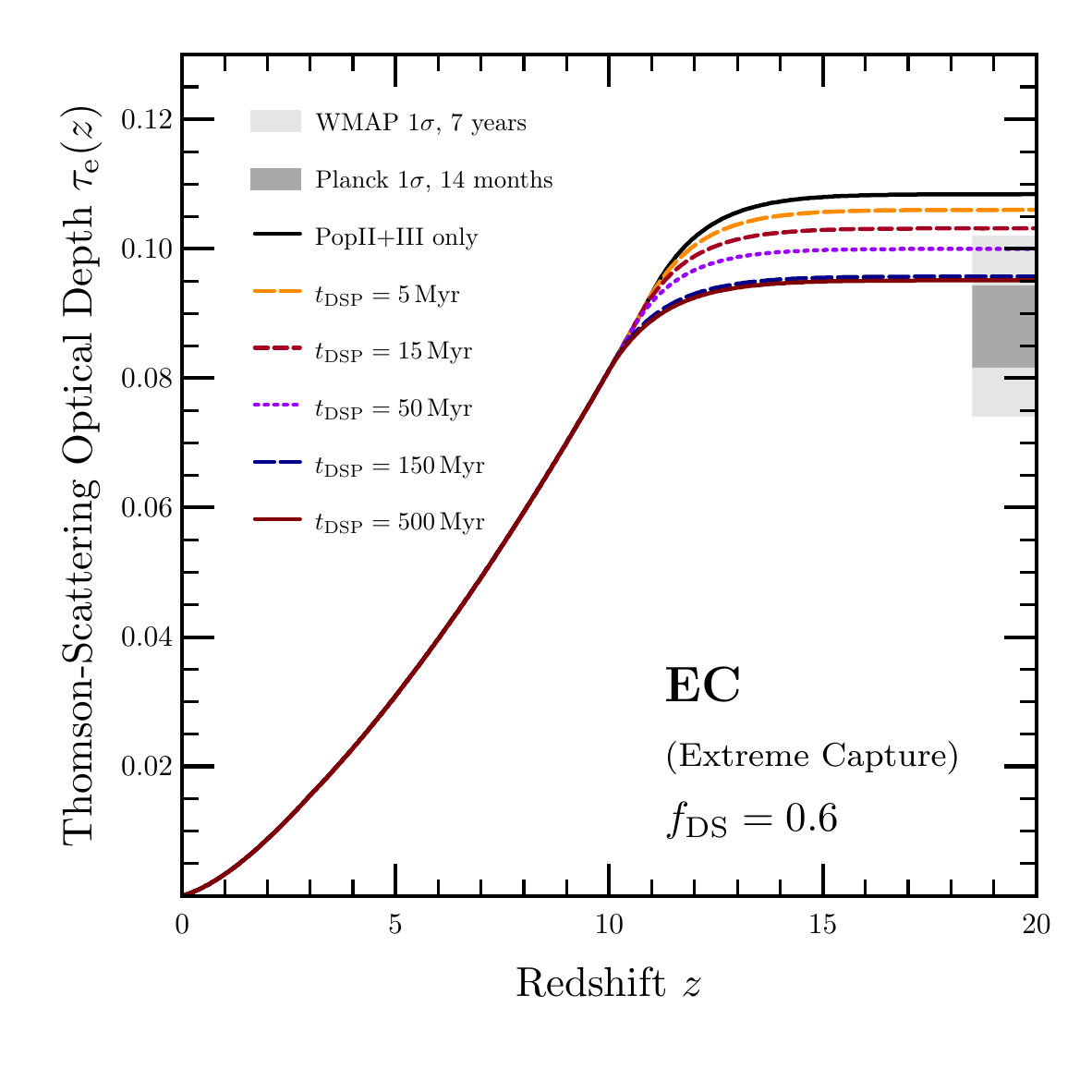}\hspace{8mm}
\includegraphics[width=0.95\columnwidth, trim = 0 20 0 0, clip=true]{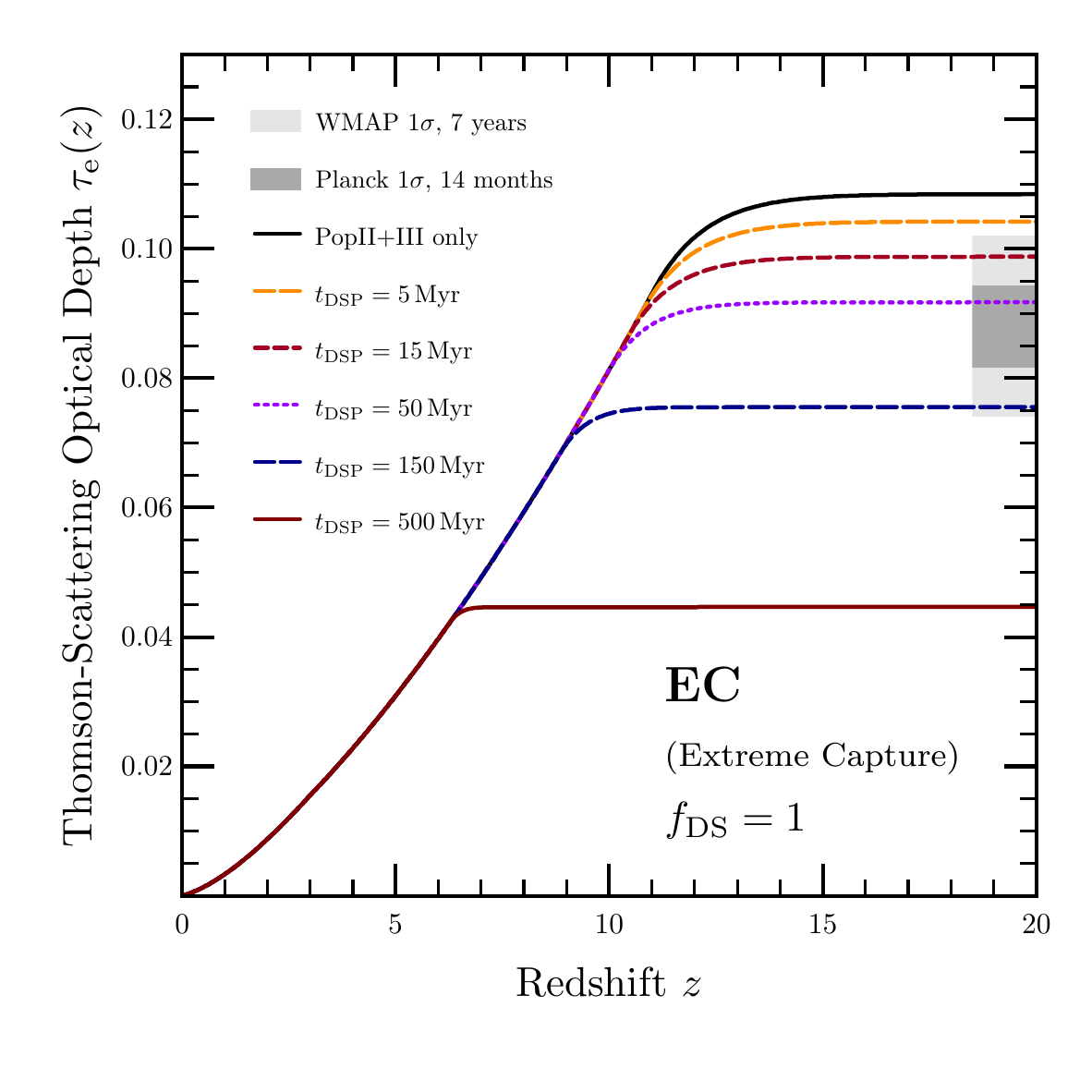}
\caption{Evolution of the lookback optical depth from the present day to redshift $z$ according to Eq.~\protect\ref{taueq}, for EC dark stars of varying lifetimes and abundances.  These curves, and the corresponding dark stellar populations, correspond to the reionization histories presented in Fig.~\ref{fig:EC_ion}.  Larger dark star fractions and more extended lifetimes reionize later, producing smaller optical depths.  For comparison, we show the WMAP7 measured $1\sigma$ band \protect\citep{WMAP7} for the optical depth to the surface of last scattering, and the corresponding error band expected from Planck \protect\citep{Colombo:09b}, assuming it measures the same central value.  Whilst these strictly correspond to a redshift $z\sim1090$, the optical depth curves have largely begun to plateau by $z\sim20$ anyway.}
\label{fig:EC_t}
\end{figure*}

\begin{figure}[tbp]
\includegraphics[width=\columnwidth]{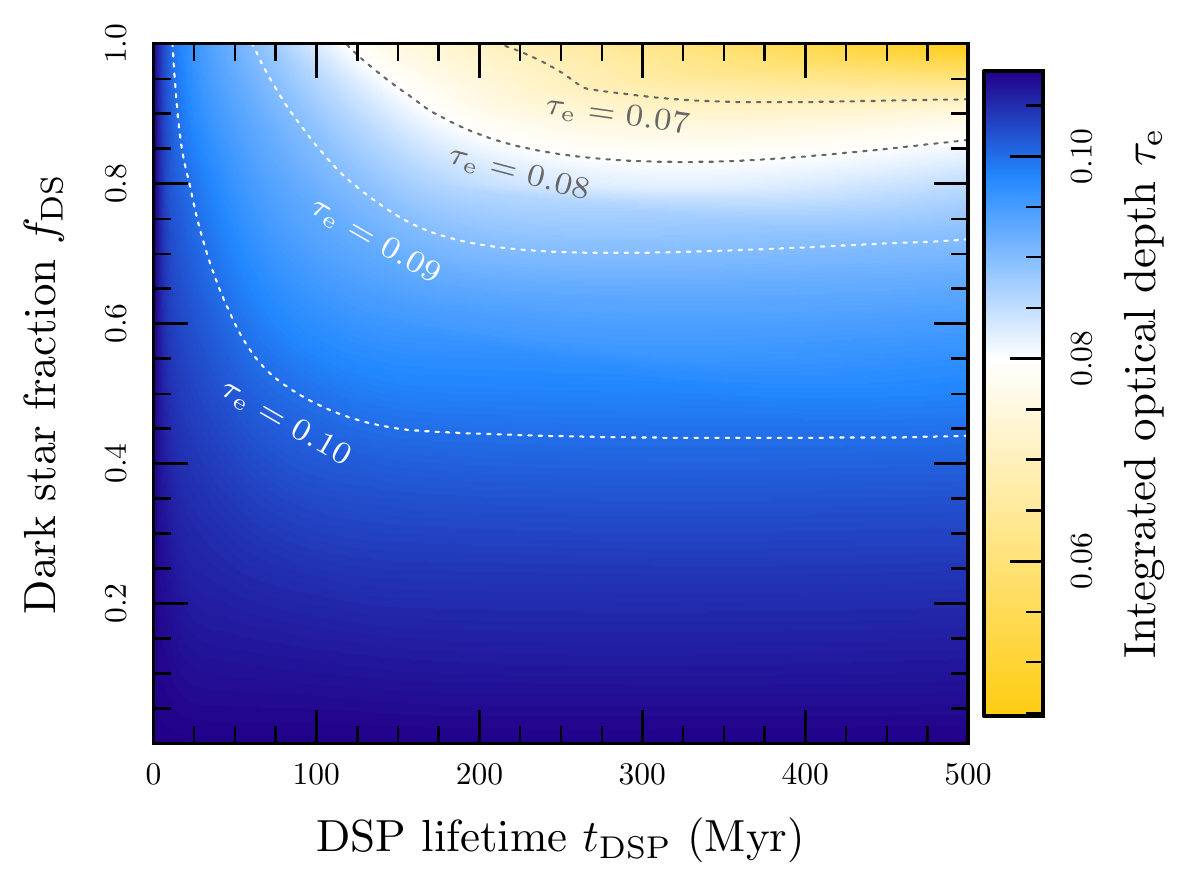}
\caption{Contours of equal integrated CMB optical depth to $z=1090$ in the EC scenario, as a function of $f_\mathrm{DS}$ and $t_\mathrm{DSP}$.  Here we performed the interpolation on the optical depths in Table~\ref{bigtable} using two-dimensional exponential tension splines \protect\citep{SRFPACK}, based on a Delauney triangulation \citep{TRIPACK} and an iterative determination of the appropriate tension factors.  Longer lifetimes and larger dark star fractions generically lead to smaller integrated optical depths; the slight upturn at large $t_\mathrm{DSP}$ in the $\tau_\mathrm{e}=0.08$ and $0.09$ contours is an artifact of the interpolation.}
\label{fig:tau_contour}
\end{figure}

\begin{figure}[tbp]
\includegraphics[width=0.95\columnwidth, trim = 0 20 0 0, clip=true]{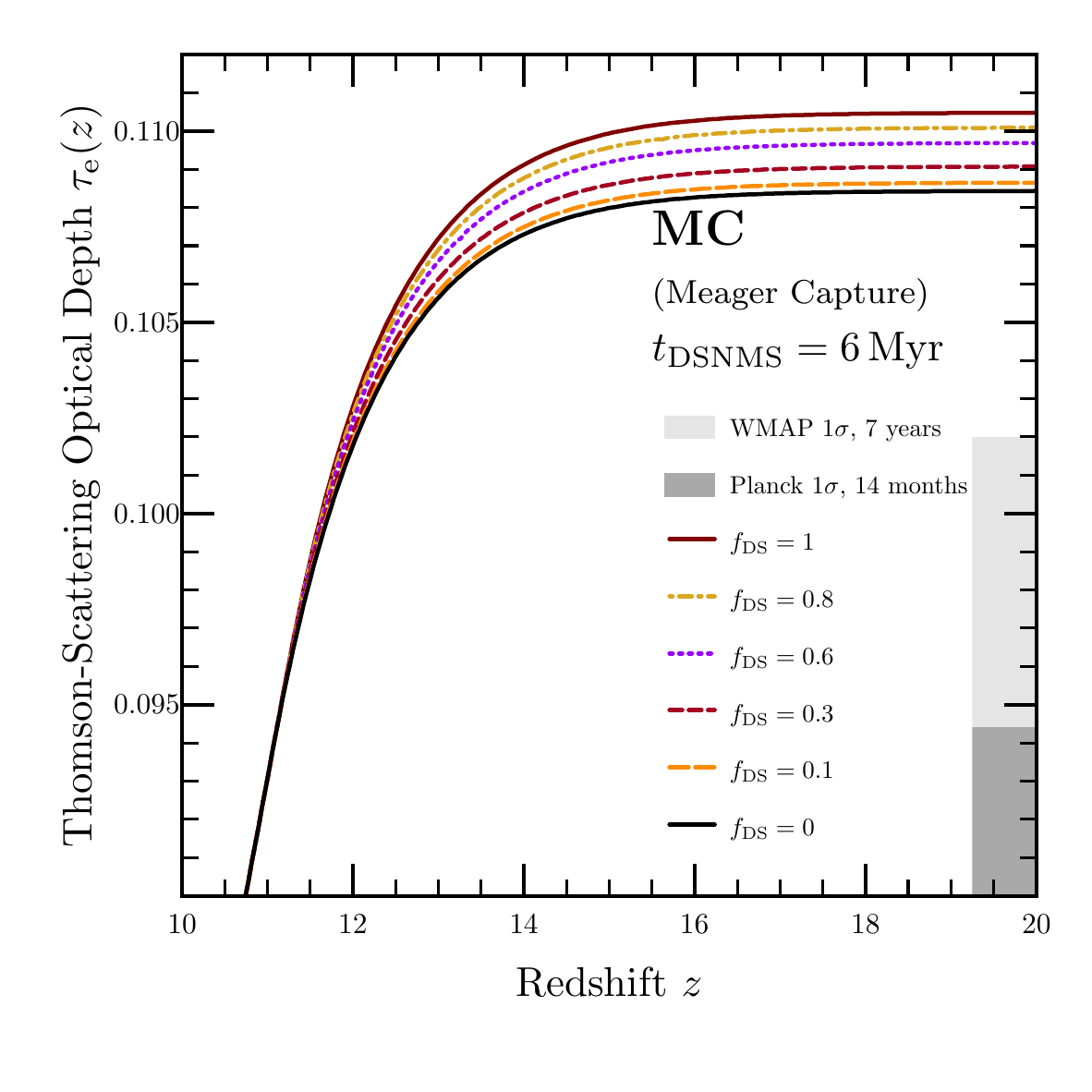}
\caption{Evolution of the lookback optical depth from the present day to redshift $z$ according to Eq.~\protect\ref{taueq}, for MC dark stars with the maximum allowed DSNMS lifetime ($t_\mathrm{DSNMS}=6$\,Myr).  These curves, and the corresponding dark stellar populations, correspond to the reionization histories presented in the right panel of Fig.~\protect\ref{fig:MC_ion}.  In the MC scenario, earlier reionization caused by increasing values of $f_\mathrm{DS}$ results in a slight increase in optical depth.  We also show WMAP7 and Planck $1\sigma$ detection/prediction bands, as per Fig.~\protect\ref{fig:EC_t}.  Note the zoomed-in axes relative to Fig.~\protect\ref{fig:EC_t}.}
\label{fig:MC_t}
\end{figure}

\begin{figure}[tbp]
\includegraphics[width=\columnwidth]{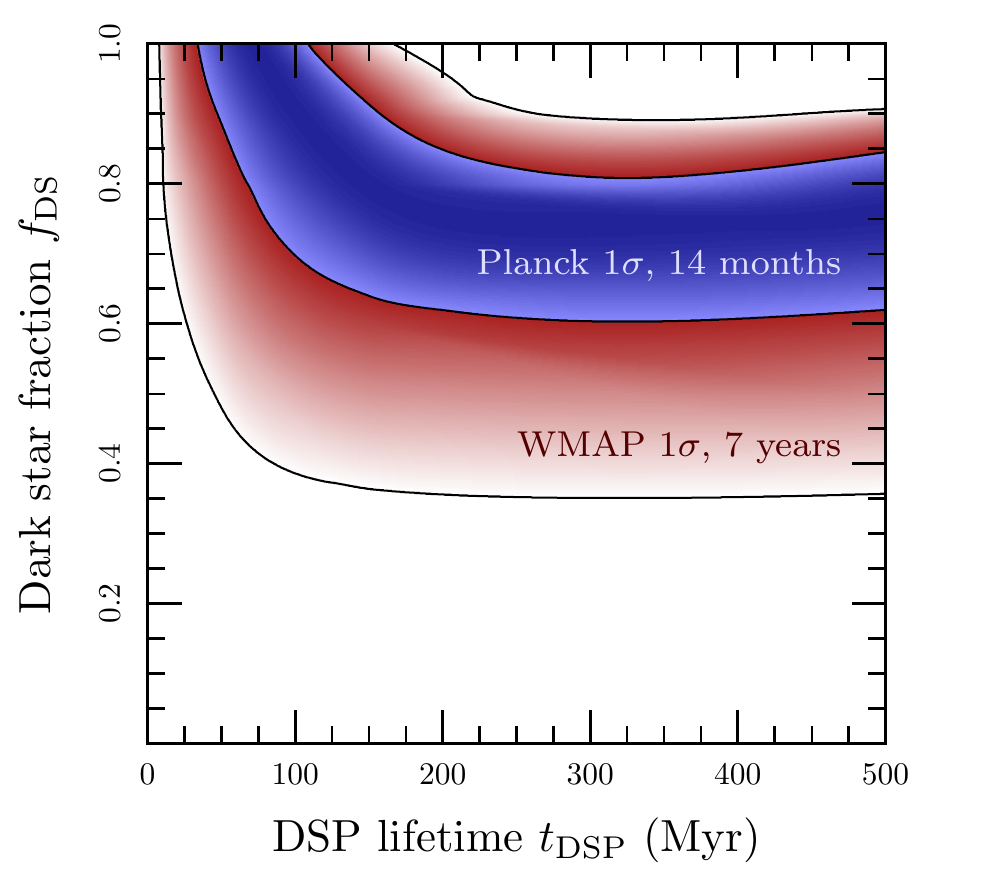}
\caption{Implied $1\sigma$ detection/exclusion regions in the $f_\mathrm{DS}$--$t_\mathrm{DSP}$ plane for EC dark stars, based on the integrated optical depth to last scattering observed by WMAP7 \protect\citep{WMAP7}. We also show projected Planck constraints \citep{Colombo:09b}, assuming that the same central value ($\tau_\mathrm{e}=0.088$) is measured by Planck as by WMAP7.  To guide the eye, the depth of shading is proportional to the likelihood of each parameter combination.  For a product $f_\star f_\mathrm{esc}=0.005$ of the star-formation efficiency and UV photon escape fraction, parameter combinations outside the red WMAP shaded region are excluded at greater than $1\sigma$ by existing data.  Combinations outside the shaded blue region will be excludable at better than $1\sigma$ by Planck.  Note however that variations in $f_\star f_\mathrm{esc}$ will shift these regions substantially; refer to discussions in the final paragraph of Sec.~\protect\ref{optdepths} and in Sec.~\protect\ref{astrouncert}.  The same interpolation methods were employed in this figure as in Fig.~\protect\ref{fig:tau_contour}; the slight upturn of the boundaries of the Planck region at large $t_\mathrm{DSP}$ is again an artifact of the interpolation.}
\label{fig:tau_constraints}
\end{figure}

\section{Impacts on the cosmic microwave background}
\label{cmb}

\subsection{Electron-scattering optical depths}
\label{optdepths}

Following \citet{Shull:08}, for each of our reionization histories we calculate the optical depth from the present day to a redshift $z$ due to Thomson scattering as
\begin{equation}
\label{taueq}
\tau_\mathrm{e}(z) = \frac{c}{H_0}\int_0^z \frac{n_\mathrm{e}(z)\sigma_\mathrm{T}}{(1+z)\left[\Omega_\mathrm{m}(1+z)^3 + \Omega_\Lambda\right]^\frac12}dz,
\end{equation}
where
\begin{equation}
n_\mathrm{e}(z) = \frac{3\Omega_\mathrm{b}H_0^2}{8\pi G m_\mathrm{H}}(1+z)^3\left[Xf_{\mathrm{H\,\textsc{ii}}}(z) +\frac{Y}{4}\left\{f_{\mathrm{He\,\textsc{ii}}}(z) + 2f_{\mathrm{He\,\textsc{iii}}}(z)\right\}\right] 
\end{equation}
is the number density of free electrons.  Here $H_0$, $\Omega_\mathrm{m}$, $\Omega_\Lambda$ and $\Omega_\mathrm{m}$ are the present-day values of the Hubble constant, mass fraction, baryon fraction and dark energy fraction of the critical density of the Universe, respectively. The electron-photon Thomson-scattering cross-section is given by $\sigma_\mathrm{T}$, the mass of hydrogen by $m_\mathrm{H}$, the primordial hydrogen mass fraction by $X$, and the primordial helium fraction by $Y\approx1-X$.  The ionization fractions $f_{\mathrm{H\,\textsc{ii}}}$, $f_{\mathrm{He\,\textsc{ii}}}$ and $f_{\mathrm{He\,\textsc{iii}}}$ refer to the fraction by number of hydrogen or helium atoms respectively in the ionization states H\,\textsc{ii}, He\,\textsc{ii} and He\,\textsc{iii}.  We assume that the number of electrons provided by ionization from He\,\textsc{i} to He\,\textsc{ii} directly tracks hydrogen ionization (i.e.~$f_{\mathrm{He\,\textsc{ii}}}+f_{\mathrm{He\,\textsc{iii}}} = f_{\mathrm{H\,\textsc{ii}}}$), leading to
\begin{equation}
n_\mathrm{e}(z) = \frac{3\Omega_\mathrm{b}H_0^2}{8\pi G m_\mathrm{H}}(1+z)^3\left[\left(1-\frac{3Y}{4}\right)f_{\mathrm{H\,\textsc{ii}}}(z) + \frac Y4f_{\mathrm{He\,\textsc{iii}}}(z)\right]. 
\end{equation}
We assume a simple step-function ionization model for $\mathrm{He\,\textsc{iii}}$, with $f_{\mathrm{He\,\textsc{iii}}}(z>3)=0$ and $f_{\mathrm{He\,\textsc{iii}}}(z\le3)=1$.  We also assume a residual electron fraction from recombination, present even before reionization at the level of $x_\mathrm{e}=2.1\times10^{-4}$.  This number comes from recombination modeling in CAMB (Sec.~\ref{pols}).  For both the optical depths based on Eq.~\ref{taueq} and CAMB calculations, we use the same values for cosmological parameters as in our reionization calculations, coming from WMAP 7-year results \citep{Larson:10}.  We give optical depths integrated up to the surface of last scattering ($z\sim1090$) for each of our parameter combinations in Table \ref{bigtable}.

In Fig.~\ref{fig:EC_t} we show the evolution of optical depth corresponding to the ionization histories detailed in Fig.~\ref{fig:EC_ion}.  As might be expected from the ionization curves, longer-lived and more numerous EC dark stars result in smaller electron-scattering optical depths, as they reionize the Universe later.  The resulting integrated optical depths across the entire EC parameter space are summarized in Fig.~\ref{fig:tau_contour}, where we plot $\tau_\mathrm{e}$ as a continuous function of $t_\mathrm{DSP}$ and $f_\mathrm{DS}$.

Similarly, we show a zoomed-in section of the optical depth curves for the longest-lived MC case ($\tau_\mathrm{DSNMS}=6$\,Myr) in Fig.~\ref{fig:MC_t}.  In this case, the smaller variations in reionization history have a correspondingly smaller (and opposite) effect upon $\tau_\mathrm{e}$, leading to slightly larger optical depths than in the $f_\mathrm{DS}=0$ case.

We also show in Figs.~\ref{fig:EC_t} and \ref{fig:MC_t} the $1\sigma$ measurement of the integrated optical depth to last scattering from WMAP7 \citep[$\tau_\mathrm{e} = 0.088\pm0.014$;][]{WMAP7}, along with a projected Planck sensitivity to the same quantity \citep{Colombo:09b}, assuming the two experiments measure the same central value.\footnote{The result we use from \protect\citet{Colombo:09b} is in fact the ratio of $1\sigma$ uncertainties on the measured value of $\tau_\mathrm{e}$ obtainable with WMAP5 and Planck in the absence of foregrounds.  We thus also assume the same percentage degradation in accuracy for both WMAP5 and Planck when mapping from expected results without foregrounds to final limits.}  Assuming that our chosen astrophysical reionization parameters in our canonical models are correct ($f_\star f_\mathrm{esc}=0.005$), a significant part of the more extreme end of the parameter space is already ruled out at better than a standard deviation by the WMAP7 measurement of $\tau_\mathrm{e}$; much more will be excluded by Planck.  This is summarized in Fig.~\ref{fig:tau_constraints}, where we plot the $1\sigma$ exclusion curves implied by WMAP7 and Planck measurements of $\tau_\mathrm{e}$ in the $t_\mathrm{DSP}$--$f_\mathrm{DS}$ plane, for the EC scenario.

Noticeably however, even the standard $f_\mathrm{DS}=0$, Pop~III+II model exhibits tension with the WMAP7 data at more than the $1\sigma$ level.  As we discuss in Sec.~\ref{astrouncert}, this should be taken with something of a grain of salt: similar variations in optical depth can be produced by reasonable changes in astrophysical parameters, so this should not be considered evidence for the existence of dark stars at this stage, even at the $1\sigma$ level.  Robust constraints could be obtained by full joint parameter scans of cosmological and (dark) reionization models.  Whilst such an exercise is well beyond the scope of this paper, the results we present in this section clearly indicate that dark stars can have a significant impact upon the reionization history of the Universe, and therefore, the integrated optical depth of the CMB.

\begin{figure*}[tbp]
\includegraphics[width=0.95\columnwidth, trim = 0 20 0 0, clip=true]{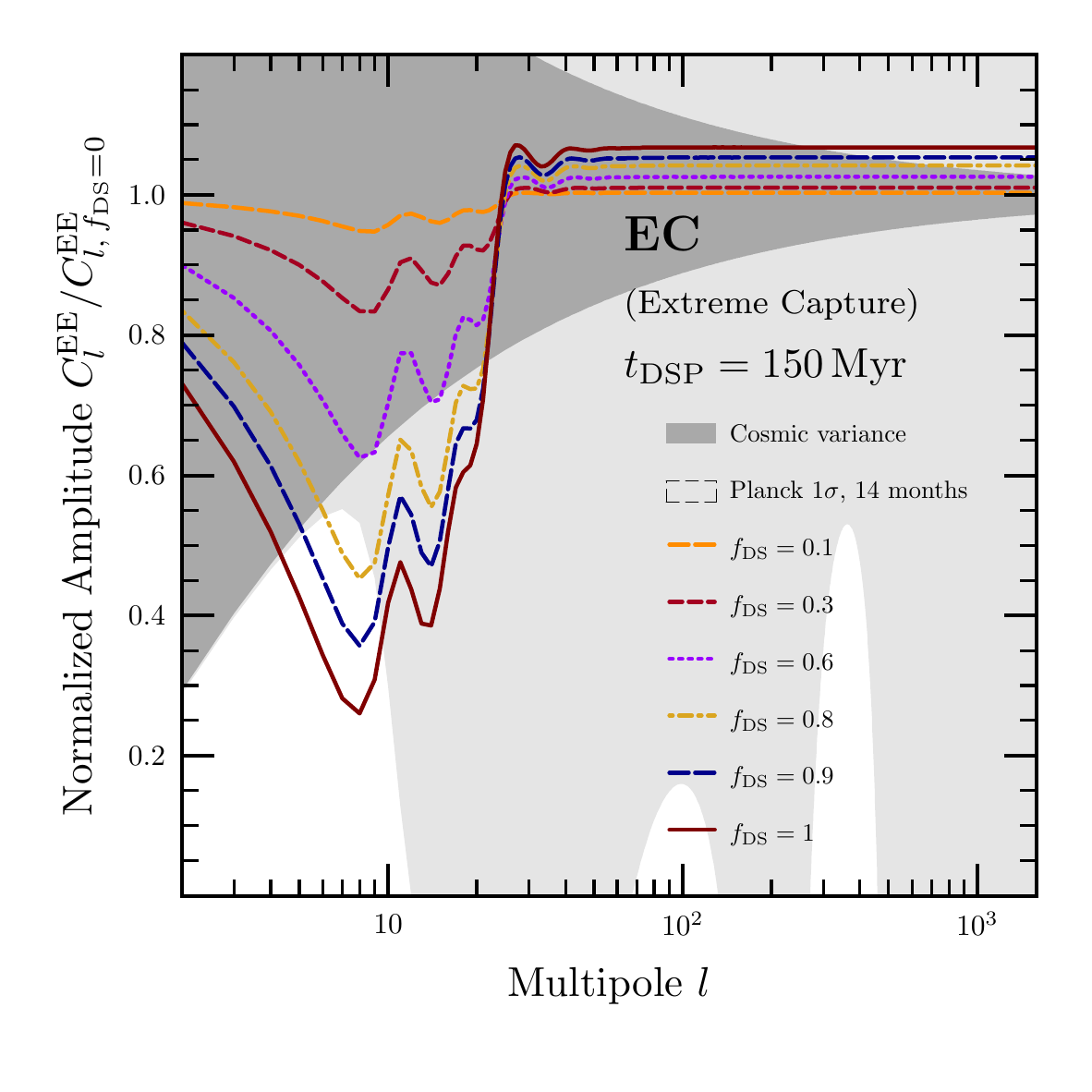}\hspace{8mm}
\includegraphics[width=0.95\columnwidth, trim = 0 20 0 0, clip=true]{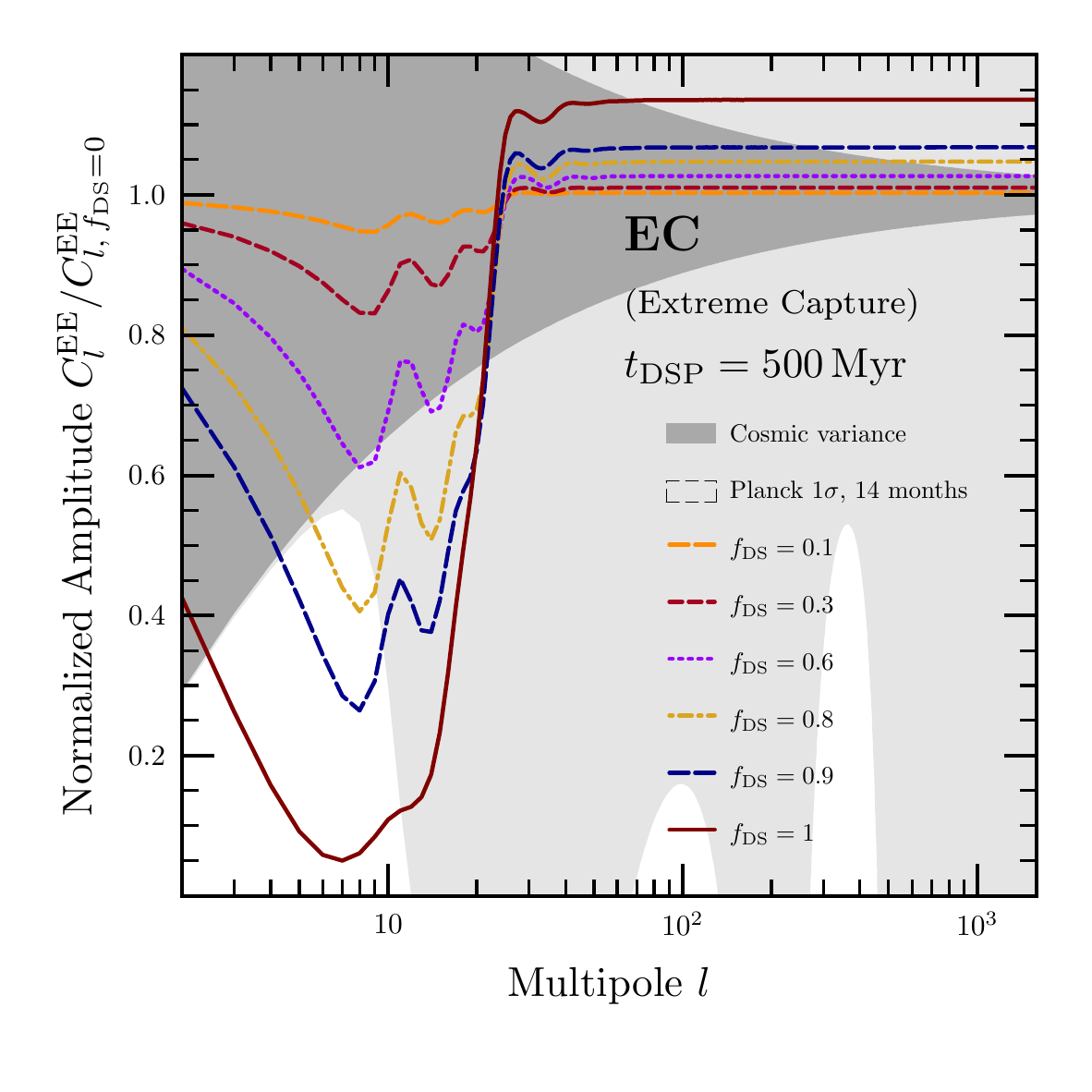}\\
\includegraphics[width=0.95\columnwidth, trim = 0 20 0 0, clip=true]{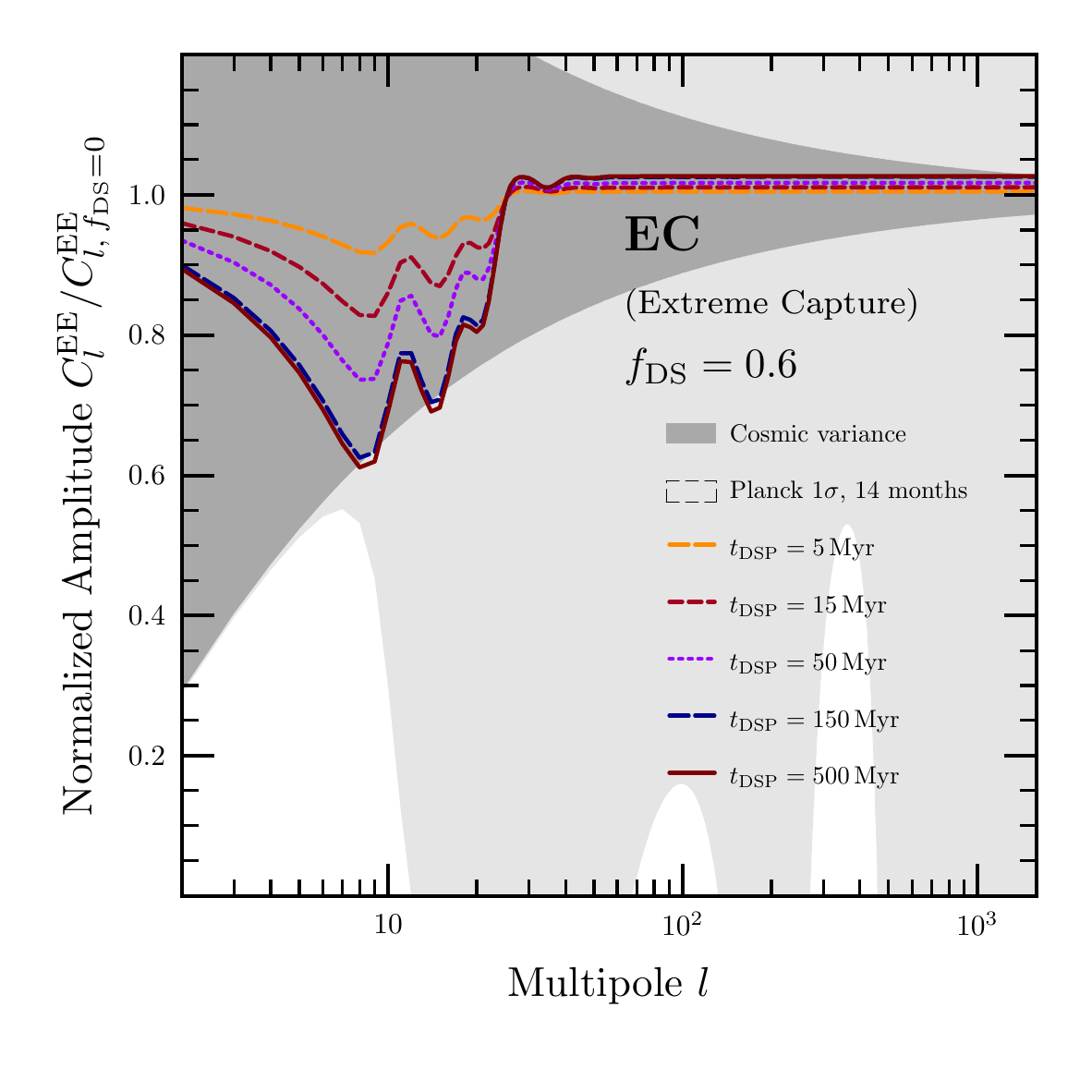}\hspace{8mm}
\includegraphics[width=0.95\columnwidth, trim = 0 20 0 0, clip=true]{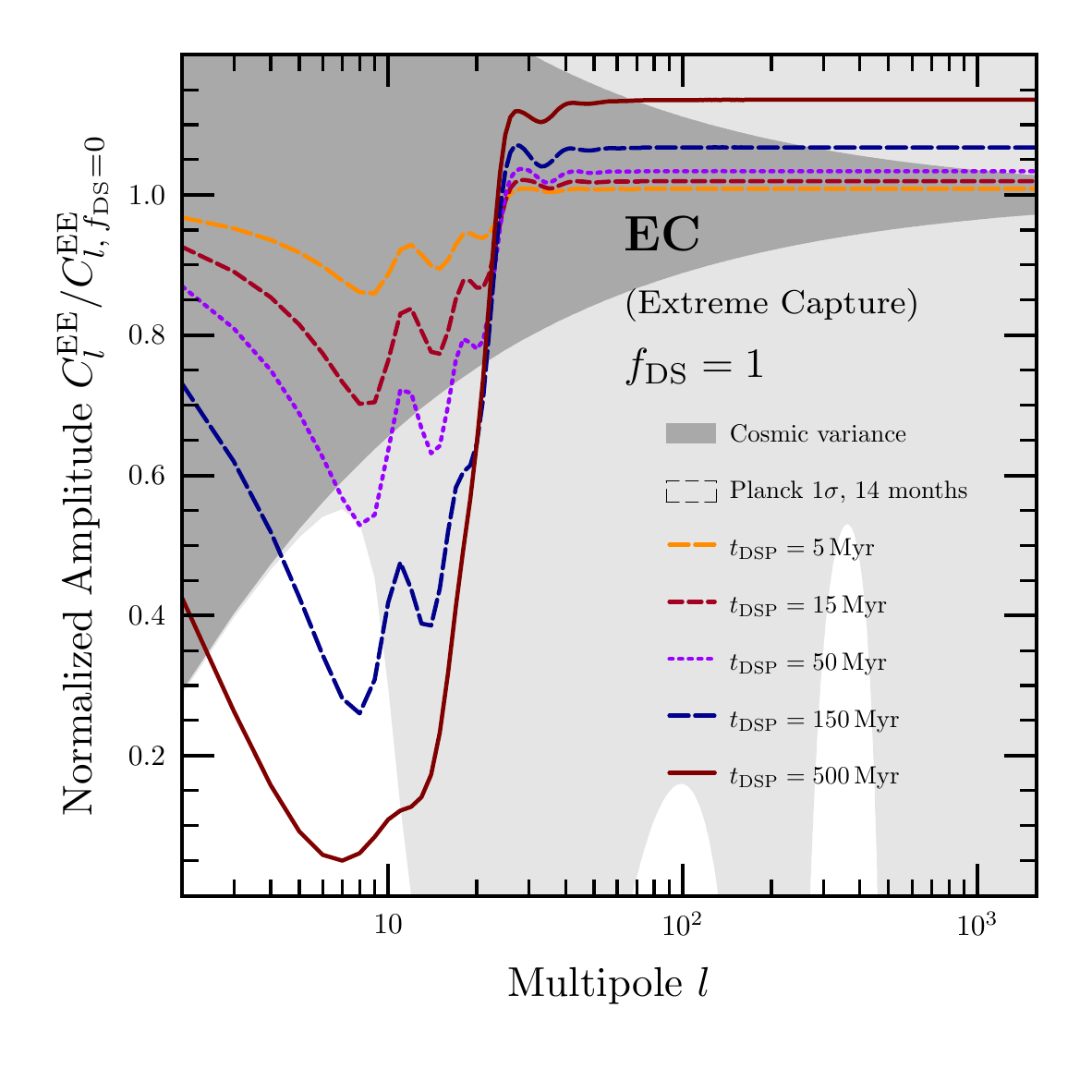}
\caption{Normalized EE CMB polarization angular power spectra, for EC dark stars of varying lifetimes and halo mass fractions.  The normalization is relative to the standard Pop~III+II case with no dark stars ($f_\mathrm{DS}=0$).  Curves and dark stellar populations correspond to the reionization histories of Fig.~\ref{fig:EC_ion}.  Larger dark star fractions and more extended lifetimes produce stronger suppressions in the EE power spectrum at large angular scales (low $l$), and stronger enhancements at small scales (large $l$).  In order to gauge the detectability of variations in the EE spectra due to dark stars, we also plot the uncertainty in the normalization due to cosmic variance \protect\citep[as per e.g.][]{Zaldarriaga:97} and the total $1\sigma$ error expected from Planck after 14 months of operation \protect\citep[cosmic variance plus combined instrumental noise in 70, 100 and 143\,GHz channels;][]{Colombo:09b}.}
\label{fig:EC_pol}
\end{figure*}

\begin{figure*}[tbp]
\includegraphics[width=0.95\columnwidth, trim = 0 20 0 0, clip=true]{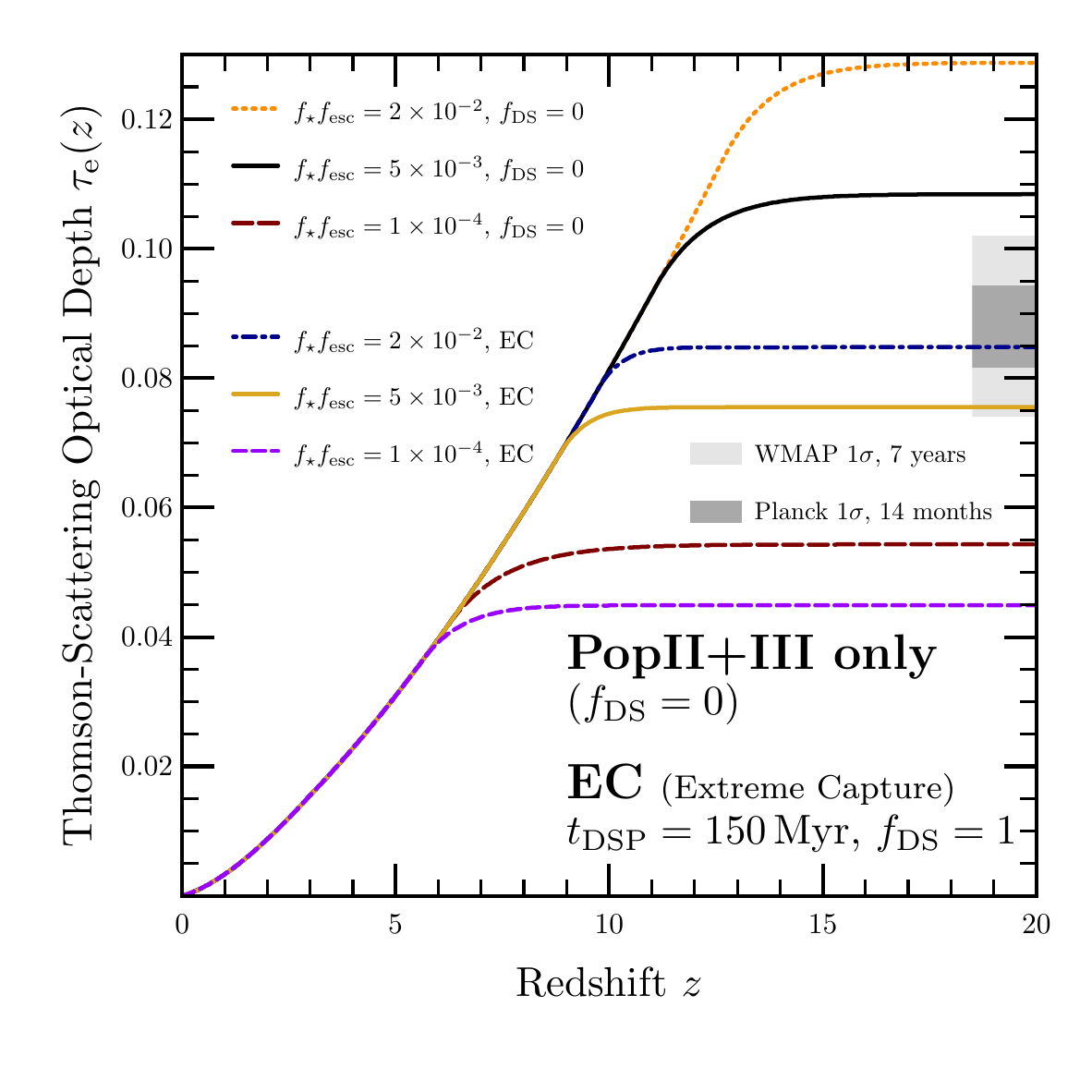}\hspace{8mm}
\includegraphics[width=0.95\columnwidth, trim = 0 20 0 0, clip=true]{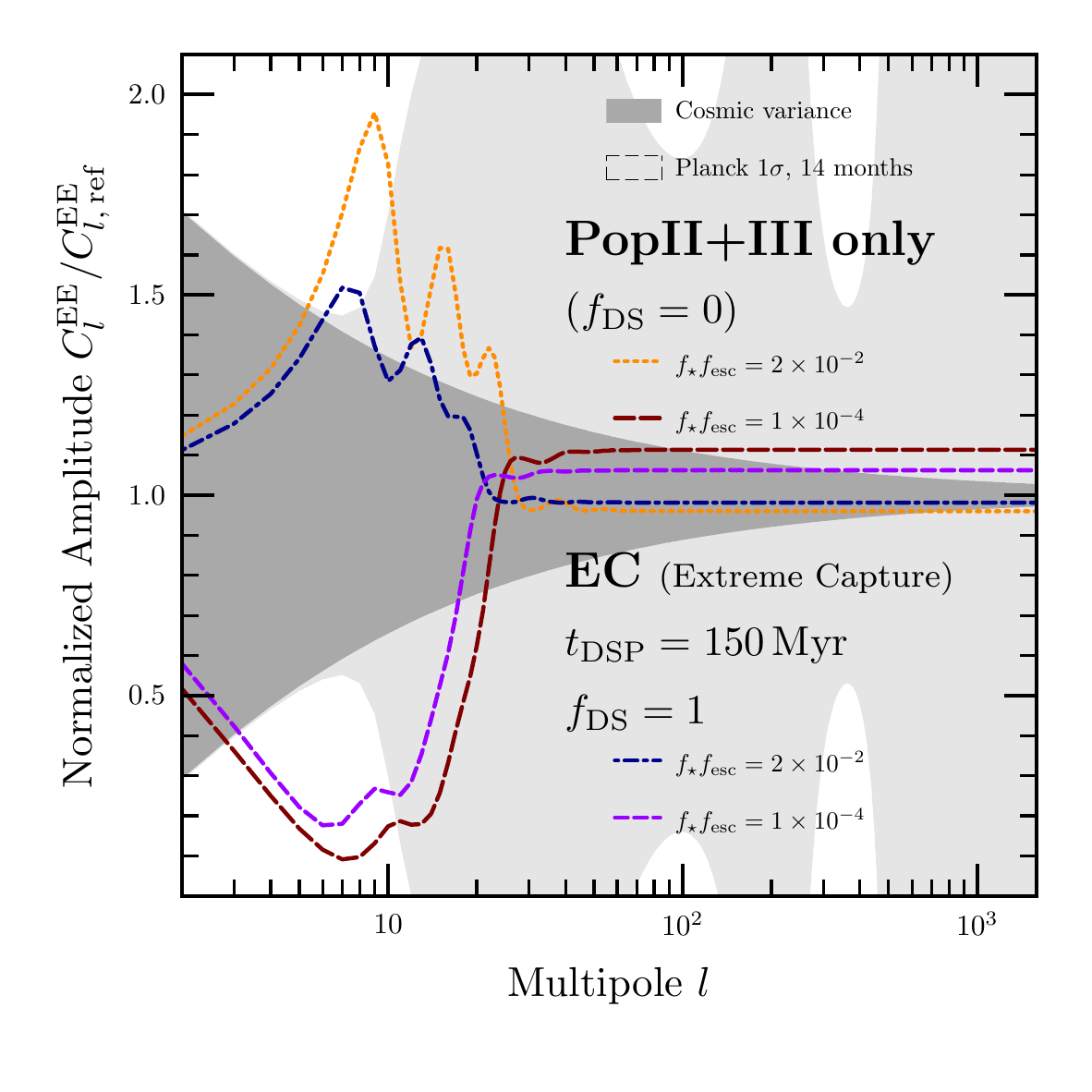}
\caption{Impacts of varying astrophysical parameters upon the evolution of electron-scattering optical depth with redshift (left) and EE polarization angular power spectra (right).  Here we again show the effects of varying the product $f_\star f_\mathrm{esc}$, both for a Universe containing no dark stars ($f_\mathrm{DS}=0$), and for one containing EC dark stars with $t_\mathrm{DSP} = 150$\,Myr, $f_\mathrm{DS}=1$.  Curves in the right panel are normalized to the corresponding EE spectra obtained with $f_\star f_\mathrm{esc}=0.005$.  The variation of astrophysical parameters can induce similar changes in optical depths and EE spectra as do EC dark stars.  Compared to EC dark stars however, which can only delay reionization, changes in $f_\star$ and $f_\mathrm{esc}$ can also speed up reionization, leading to larger optical depths and enhanced large-scale (low $l$) power in the EE spectrum.}
\label{fig:AST_cmb}
\end{figure*}
\subsection{Polarization}
\label{pols}

We calculate the effects of the different reionization histories on the polarization (EE) CMB power spectra by modifying the Boltzmann code CAMB\footnote{\href{http://camb.info}{http://camb.info}} \citep{Lewis:00}.  Instead of the simple hydrogen reionization model included in CAMB, our modified version uses the H ionization fractions presented in Sec.~\ref{reionhist} as the basis for its reionization calculations.  We include contributions to the total electron fraction from H~\textsc{ii}, He~\textsc{ii} and He~\textsc{iii} with the assumptions stated earlier, as well as a residual electron fraction.  As in the standard CAMB reionization calculation, we assume that electrons from He~\textsc{ii} track those from H~\textsc{ii}, and model contributions from He~\textsc{iii} with a smoothed step function centered at $z\sim 3.5$.  We take the residual electron fraction after recombination to be $x_\mathrm{e}=2.1\times10^{-4}$, based on the output of RECFAST within CAMB.  Using CAMB's highest accuracy setting, we calculated the temperature (TT), polarization (EE) and cross (TE) power spectra, as well as integrated optical depths.  We checked that the optical depths computed with CAMB agree to within their stated numerical accuracy with those from Eq.~\ref{taueq} (as presented in Table~\ref{bigtable}), and verified that the slight difference in the treatment of He~\textsc{iii} here and in Sec.~\ref{optdepths} has a negligible impact upon integrated optical depths.

In Fig.~\ref{fig:EC_pol} we plot EE polarization spectra for the same EC cases as illustrated in Figs.~\ref{fig:EC_ion} and \ref{fig:EC_t}.  The spectra are normalized to the corresponding EE spectrum obtained in the standard $f_\mathrm{DS}=0$, Pop~III+II scenario, in order to investigate the ability of CMB experiments to distinguish dark stars from standard reionization.  To this end, we also plot the uncertainty on the normalization due to cosmic variance, and the combination of cosmic variance and total readout noise expected across the 70, 100 and 143\,GHz channels in the first 14 months of Planck operation \citep{Colombo:09b}.  Large parts of the parameter space are distinguishable from $f_\mathrm{DS}=0$ in a cosmic-variance-limited experiment, and a number of the more extreme scenarios are even detectable by Planck at better than $1\sigma$.  Although essentially all such models may be disfavored anyway by Planck's measurement of the integrated optical depth, the EE spectrum would nonetheless provide a complementary (albeit weak) statistical handle via which to increase the power of full parameter scans to exclude such dark star models.

We do not show TT or TE spectra for the EC scenario, as they exhibit less striking deviations from the corresponding spectra of the standard Pop~III+II scenario at low multipoles $l$ (large angular scales) than the EE curves do.  We do point out however that the TT and TE spectra exhibit damping at large $l$ due to the changing optical depth, which is more clearly visible than in the EE spectra.  We also do not show power spectra for the MC or NC cases, as they show little deviation in general from the standard Pop~III+II case.

\subsection{Astrophysical uncertainties and implications for parameters of reionization models}
\label{astrouncert}

In Fig.~\ref{fig:AST_cmb} we show the impacts of varying astrophysical parameters upon CMB observables.  Here we give the variations in optical depth and EE polarization resulting from the same variations of $f_\star f_\mathrm{esc}$ as in Fig.~\ref{fig:AST_ion}.  For EE spectra, we normalize to the corresponding curve with default astrophysical parameters in each case; i.e. to the standard $f_\star f_\mathrm{esc}=0.005$ Pop~III+II spectrum for the $f_\mathrm{DS}=0$ curves, and to the $f_\mathrm{DS}=1$, $t_\mathrm{DSP}=150$\,Myr, $f_\star f_\mathrm{esc}=0.005$ spectrum for the curves where $f_\mathrm{DS}=1$, $t_\mathrm{DSP}=150$\,Myr.  Comparing with Figs.~\ref{fig:EC_t} and \ref{fig:EC_pol}, we see again that variations in the astrophysical parameters within reasonable ranges can have similar effects (both in strength and character) to variations in the dark star parameters.  Although this means that the impact of dark stars on the CMB is very difficult to unambiguously disentangle from existing theoretical uncertainties in reionization modeling, it also serves as a clear indication that the potential effect of dark stars upon CMB observables affected by reionization could be quite significant.

In some cases, it is possible that specific regions of the reionization parameter space that are ruled out in standard Pop~III+II-only scenarios by the current WMAP7 (or projected Planck) data, are reopened by the possibility of having dark stars. For example, Fig.~\ref{fig:AST_cmb} reveals that the extreme cases of varying astrophysical parameters ($f_\star f_\mathrm{esc} = 0.02$ or $10^{-4}$) are ruled out for the Pop III+II-only scenarios by current CMB data. However, the case of $f_\star f_\mathrm{esc}$ = 0.02 is {\it not} ruled out for the EC dark star case with $f_\mathrm{DS}=1$, $t_\mathrm{DSP}=150$\,Myr. Having more dark stars in the EC scenarios mimics a reduction in the astrophysical efficiency -- i.e.~increasing $f_{\rm DS}$ allows for {\it greater} astrophysical efficiency (larger $f_\star$, larger $f_\mathrm{esc}$ or both) in reionization models. This is potentially an important factor to consider when placing constraints on the astrophysical aspects of reionization from CMB data \citep{Haiman:03, Shull:08}.

The impacts of uncertainties in cosmological parameters on $\tau_\mathrm{e}$ are subdominant to those from astrophysical uncertainties. For the same two scenarios shown in Fig.~\ref{fig:AST_cmb} (no dark stars and EC dark stars with $f_\mathrm{DS}=1$ and $t_\mathrm{DSP}=150$\,Myr), we find that varying $\sigma_8$ over the 1$\sigma$ allowed range of the WMAP7 fit we used as input for all calculations ($\sigma_8 = 0.801\pm0.030$), leads to a change in the redshift of reionization of less than 1, and a change of no more than 0.01 in $\tau_e$.  This is relatively small compared with the variations in Fig.~\ref{fig:AST_cmb}. 

\section{Conclusions}
\label{conclusion}

We have calculated reionization histories, CMB optical depths and anisotropy power spectra for a broad range of stellar population models containing dark stars.  We identified three distinct regimes: no capture (NC), where dark stars live as cool, semi-diffuse objects for no more than $\sim 0.4$\,Myr and then quickly move on to the main sequence; meager capture (MC), where dark stars undergo the same initial evolution as in the NC case, but exhibit slightly decreased surface temperatures and moderately increased lifetimes on the main sequence, and extreme capture (EC), where dark stars live an extended life as cool, semi-diffuse objects before reaching the main sequence.

NC dark stars have effectively no impact on reionization or its signatures in the CMB.  MC dark stars cause the Universe to reionize more quickly the longer-lived and more numerous they are, producing a slight increase in the redshift of reionization.  This leads to small increases in the total integrated optical depth of the CMB.  

EC dark stars can have dramatic and completely opposite effects to MC dark stars, delaying reionization to as late as $z\sim 6$.  Greater numbers of EC dark stars and longer dark star lifetimes lead to increased delays, and correspondingly decreased integrated optical depths to the last scattering surface of the CMB.  Using WMAP7 observations of the integrated optical depth, we have been able to rule out more extreme areas of the parameter space covered by the EC scenario.  Planck will improve these bounds significantly.  EC dark stars also produce a characteristic suppression of EE polarization power on the largest scales of the CMB, as well as a slight enhancement at small scales due to the decreased optical depth.  

Many of these effects can be mimicked or compensated for by changes in the astrophysical parameters of standard reionization models.  Disentangling the impact of dark stars from other theoretical uncertainties in reionization modeling will be challenging, even with Planck.  Not only can dark stars have a substantial impact on reionization and its signatures in the CMB, but the addition of dark stars to standard reionization models can in fact substantially increase the range of astrophysical parameters that can be made consistent with existing (and future) observations.

\vspace{3mm}
\acknowledgments
P.S. thanks the organizers of the Cosmic Radiation Fields 2010 Workshop at DESY Hamburg, for providing a stimulating environment for discussion of an early version of these results, Dominik Schleicher and Erik Zackrisson in particular for such discussions, and the Department of Physics and Astronomy at the University of San Francisco, the Department of Physics at Stockholm University and the Max Planck Institute for Astrophysics for their hospitality whilst this work was being completed.  P.S. is supported by the Lorne Trottier Chair in Astrophysics and an Institute for Particle Physics Theory Fellowship.  A.V. gratefully acknowledges support from Research Corporation through the Single Investigator Cottrell College Science Award, and from the University of San Francisco Faculty Development Fund.  The work of P.G. was supported in part by NSF award PHY-0456825 and NASA contract NNX09AT70G.  E.P. thanks Loris Colombo for useful discussions, and acknowledges support from NASA grant NNX07AH59G and JPL Planck subcontract 1290790.  G.H. acknowledges support from a Canada Research Chair, NSERC, and the Canadian Institute for Advanced Research.

\bibliography{DMbiblio,DSPaper-extra}

\end{document}